\documentclass[aps,prx,twocolumn,superscriptaddress,showpacs,floatfix,longbibliography]{revtex4-1}
\usepackage{amsmath,amssymb,graphicx}

\usepackage[utf8]{inputenc}
\usepackage[T1]{fontenc}
\usepackage{xcolor}
\usepackage{dsfont}

\IfFileExists{newtxtext.sty}
{\usepackage{newtxtext,newtxmath}}
{\IfFileExists{stix.sty}
	{\usepackage{stix}}
	{\IfFileExists{mathptmx.sty}
		{\usepackage{mathptmx}}{} } }

\usepackage{textcomp}

\usepackage{bm}

\IfFileExists{siunitx.sty}{\usepackage{booktabs,siunitx}}{}

\pdfoutput=1
\usepackage{color}
\definecolor{LinkColor}{rgb}{0.256,0.439,0.588}
\usepackage{hyperref}
\hypersetup{
	pdfauthor={Wei Wang},
	pdftitle={paper},
	colorlinks=true,
	citecolor=LinkColor,
	linkcolor=LinkColor,
	urlcolor=LinkColor
}

\renewcommand{\vec}[1]{\mathbf{#1}}
\newcommand{\vect}[1]{{\bm{#1}}}
\newcommand{\bra}[1]{\langle#1\rvert}
\newcommand{\ket}[1]{\lvert#1\rangle}

\newcommand{\ii}{\mathrm{i}}
\usepackage{pifont}

\newcommand{\Tr}{\mathop{\mathrm{Tr}}}

\renewcommand{\Im}{\operatorname{Im}}
\newcommand{\U}{\mathrm{U}}
\newcommand{\SU}{\mathrm{SU}}
\newcommand{\PSU}{\mathrm{PSU}}

\newcommand{\eqnref}[1]{Eq.\,\eqref{#1}}

\newcommand{\tabref}[1]{Tab.\,\ref{#1}}

\begin{document}
\title{Dynamics of Compact Quantum Electrodynamics at Large Fermion Flavor}
\author{Wei Wang}
\affiliation{Beijing National Laboratory for Condensed Matter Physics and Institute of Physics,Chinese Academy of Sciences, Beijing 100190, China}
\affiliation{School of Physical Sciences, University of Chinese Academy of Sciences, Beijing 100190, China}
\author{Da-Chuan Lu}
\affiliation{Department of Physics, University of California at San Diego, La Jolla, CA 92093, USA}
\author{Xiao Yan Xu}
\affiliation{Department of Physics, Hong Kong University of Science and Technology, Clear Water Bay, Hong Kong, China}
\author{Yi-Zhuang You}
\affiliation{Department of Physics, University of California at San Diego, La Jolla, CA 92093, USA}
\author{Zi Yang Meng}
\affiliation{Beijing National Laboratory for Condensed Matter Physics and Institute of Physics,Chinese Academy of Sciences, Beijing 100190, China}
\affiliation{Department of Physics, The University of Hong Kong, China}
\affiliation{CAS Center of Excellence in Topological Quantum Computation and School of Physical Sciences, University of Chinese Academy of Sciences, Beijing 100190, China}
\affiliation{Songshan Lake Materials Laboratory, Dongguan, Guangdong 523808, China}

\begin{abstract}
 Thanks to the development in quantum Monte Carlo technique, the compact U(1) lattice gauge theory coupled to fermionic matter at (2+1)D is now accessible with large-scale numerical simulations, and the ground state phase diagram as a function of fermion flavor ($N_f$) and the strength of gauge fluctuations is mapped out~\cite{Xiao2018Monte}. Here we focus on the large fermion flavor case ($N_f=8$) to investigate the dynamic properties across the deconfinement-to-confinement phase transition. In the deconfined phase, fermions coupled to the fluctuating gauge field to form U(1) spin liquid with continua in both spin and dimer spectral functions, and in the confined phase fermions are gapped out into valence bond solid phase with translational symmetry breaking and gapped spectra. The dynamical behaviors provide supporting evidence for the existence of the U(1) deconfined phase and could shine light on the nature of the U(1)-to-VBS phase transition which is of the QED$_3$-Gross-Neveu chiral O(2) universality whose properties still largely unknown.
\end{abstract}

\maketitle

\section{Introduction}
For several decades, the topic of dynamical coupling between lattice gauge fields and fermionic matter fields have attracted considerable attention among physicists from high-energy~\cite{PhysRevD.42.3520,PhysRevD.84.014502,PhysRevD.94.065026,PhysRevD.97.054510,PhysRevLett.121.041602, PhysRevD.90.036002, PhysRevD.94.056009} and condensed matter~\cite{PhysRevB.37.3774,PhysRevB.39.11538,PhysRevB.57.6003,PhysRevLett.86.3871,PhysRevB.66.144501,PhysRevB.62.7850,PhysRevB.66.205104,PhysRevLett.91.171601,PhysRevB.70.214437,PhysRevB.71.075103,PhysRevB.72.104404,PhysRevB.76.149906,PhysRevB.77.045107,PhysRevX.7.031020,PhysRevX.6.041049,gazit2017emergent,gazit2018confinement,PhysRevB.96.205104} communities. Previous works have quickly established the understanding at the large fermion flavor $(N_f)$ limit~\cite{PhysRevB.37.3774,PhysRevB.39.11538,PhysRevB.57.6003,PhysRevLett.86.3871,PhysRevB.66.144501,PhysRevB.62.7850,PhysRevB.66.205104,PhysRevLett.91.171601,PhysRevB.70.214437}, as $1/N_f$ expansion is controlled thence, but left the physically most interesting cases of small $N_f$ -- for example $N_f=2$ corresponds to the spin-1/2 case of electrons -- unsolved. The very recent breakthrough of quantum Monte Carlo (QMC) simulations of Z2~\cite{PhysRevX.6.041049,gazit2017emergent,gazit2018confinement,ChuangChen2019} and U(1)~\cite{Xiao2018Monte} gauge fields coupled to fermions provide the possibility of concrete investigations at the small $N_f$, and the expected deconfinement-to-confinement phase transitions and special properties of these phases are discovered. In such settings, the interactions between fermions are mediated via the fluctuating gauge bosons, which resemble the situation of fractionalized particles and emergent gauge fields in several prototypical strongly correlated systems including, but not limited to, the low-energy description of the high-temperature superconductors~\cite{PhysRevB.57.6003,PhysRevB.66.205104,PhysRevB.72.104404}, frustrated magnets~\cite{PhysRevB.72.104404,PhysRevB.65.224412,PhysRevB.69.064404} and deconfined quantum criticalities~\cite{PhysRevB.70.144407, PhysRevLett.98.227202,PhysRevX.7.031052,PhysRevB.98.174421,Li2019Deconfined,Zhou2019Quantum}.  The quantum phases and phase transitions discovered are clearly beyond the Landau-Ginzburg-Wilson
paradigm built upon the concepts of symmetry-breaking and local order parameters, and served as the building bulks of the new paradigm of quantum matter.

As for the discrete Z2 gauge field coupled to
fermionic matter in (2+1)D~\cite{PhysRevX.6.041049,gazit2017emergent,gazit2018confinement}, the deconfined phase with fractionalized fermionic excitations at weak gauge fluctuation and confined phase with symmetry-breaking at strong gauge fluctuation have been revealed. The (Z2) deconfinement-to-confinement transitions are continuous and associated with fermion gap opening in the excitation spectrum. Further developments that involve not only Z2 gauge but also Z2 matter fields to dynamically couple to Fermi surface (FS) give rise to the long-thought orthogonal metal phase which has metallic transport but no quasiparticles at the FS~\cite{ChuangChen2019}, probably the simplest non-Fermi-liquid that can be generated without ambiguity~\cite{PhysRevB.86.045128} in (2+1)D lattice models.

As for the continuous U(1) gauge field coupled to fermionic matter at (2+1)D, such as the compact quantum electrodynamics (cQED$_3$), there are fundamental physical questions awaiting for affirmative answer. The pure gauge theory at (2+1)D is known to be always confined~\cite{PhysRevLett.91.171601,polyakov1977quark,mandelstam1976vortices,case2004self}, but whether the coupling to gapless fermionic
matter could drive the system towards deconfinement have been debated~\cite{PhysRevLett.91.171601,polyakov1977quark,mandelstam1976vortices,case2004self,XYSong2018}. As mentioned above, large $N_f$ limits~\cite{PhysRevB.70.214437,PhysRevB.76.149906,PhysRevB.77.045107,Unsal2008} demonstrated the existence of the U(1) deconfined phase, but previous QMC works at medium and small $N_f$ values are shown inconclusive~\cite{PhysRevD.84.014502,PhysRevD.94.065026,PhysRevD.97.054510,PhysRevLett.121.041602} due to the difficulties in effectively simulating the continuous gauge fields in the $(2+1)$ space-time with zero modes at the fermion spectra. It is only till very recent, that in Ref.~\cite{Xiao2018Monte}, with the help of fast updates and high level parallelization, that the phase diagram of U(1) gauge field coupled to fermion field at small $N_f$ has been mapped out, and the existence of U(1) deconfined phase, or algebraic quantum spin liquid in the condensed matter parlance~\cite{PhysRevB.72.104404}, for $N_f=2,4,6,8$ are discovered with certainty. However, even the latest QMC simulations are still by no means easy, suffering long autocorrelation time in the critical phase of U(1) deconfined spin liquid, the transitions from U(1) deconfined phase to various confined phases (antiferromagnetic insulator (AFM) phase for $N_f=2$, and valence bond solid (VBS) phase and AFM for $N_f=4,6,8$) have not been able to investigated in detail, both statically and dynamically.

\begin{figure*}
\includegraphics[width =\textwidth]{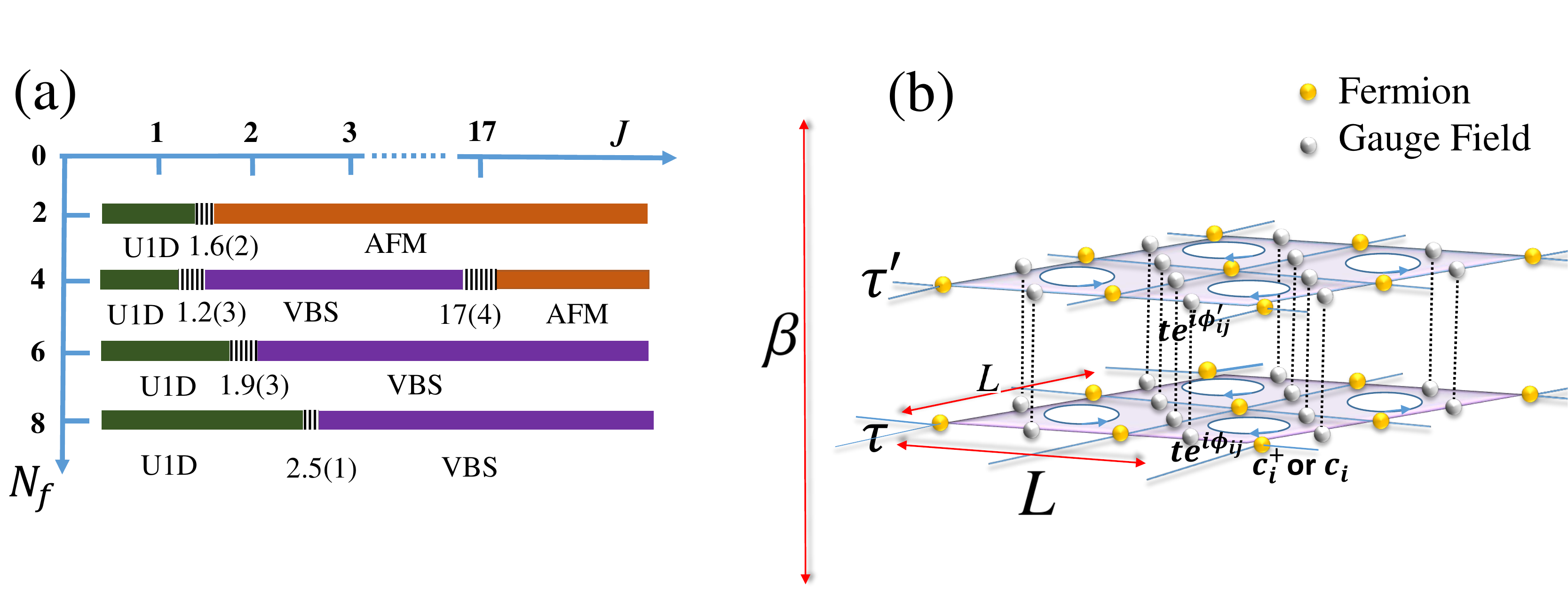}
\caption{(a) Phase diagram spanned by the fermi flavors $N_f$ and the strength of gauge field fluctuations $J$ of the quantum rotor model in Eq.~\eqref{eq:eq1}. U1D stands for the U(1) deconfined phase, AFM stands for the antiferromagnetic Mott insulator phase and VBS stands for the valence bond solid phase. The diagram and the values of critical points are adapted from Ref.~\cite{Xiao2018Monte}. (b) The illustration of the quantum rotor model in Eq.(\ref{eq:eq1}). $L$ and $\beta=1/T$ are the space-time dimensions of the lattice model. The yellow balls represent fermions and the white balls represent the gauge field attached to the nearest-neighbor fermion hopping. The blue circle lines stand for the flux term per plaquette. The gauge fields fluctuate from $\phi_{ij}$ at imaginary time slice $\tau$ to $\phi'_{ij}$ at time slice $\tau'$.}
\label{fig:fig1}
\end{figure*}

These transitions, dubbed QED$_3$-Gross-Neveu chiral Heisenberg (QED$_3$-GN-O(3)) or XY (QED$_3$-GN-O(2)) transitions, are of high interests to both condensed matter and high-energy physicists, as the phase transition of algebraic quantum spin liquid to other magnetically order phases have experimental relevance. Inspired by the numerical work of Ref.~\cite{Xiao2018Monte}, there are recently several analytical works addressing the critical properties of them~\cite{JGracey2018,Ihrig2018,Zerf2019,Dupuis2019,PhysRevB.98.165125,PhysRevB.99.195135}. And the conclusions drawn there are that the U(1)-to-AFM and U(1)-to-VBS phase transitions are indeed possible and the higher-order perturbative RG calculations performed also suggest the possible range of critical exponents of these QED$_3$-GN transitions within $1/N_f$ and $4-\epsilon$ expansions~\cite{JGracey2018,Ihrig2018,Zerf2019,Dupuis2019,PhysRevB.98.165125,PhysRevB.99.195135}.

While the QMC evaluation of the critical exponents are still difficult (currently the largest system accessed are $L=20$ due to the aforementioned computational complexity), the dynamical signatures of the transition would then provide guiding evidence for comprehensive understanding of them. Similar as the case of deconfined quantum critical point with emergent O(4) symmetry, where the coupling effects of fractionalized spinon and emergent U(1) gauge fields manifest in the spin spectral functions~\cite{PhysRevB.98.174421}, the unearthness of the QED$_3$-GN dynamical signatures will provide similar physical understanding. In terms of quantum Monte Carlo simulations, the dynamical signatures can be obtained in two steps. One first measures the imaginary time correlation functions with good statistics, then performs analytic continuation to convert the correlation from imaginary to real frequencies. The recent developments of stochastic analytic continuation (SAC) scheme~\cite{PhysRevE.94.063308} is proven to be more reliable and could reveal non-trivial results in both unfrustrated and frustrated magnetic systems in 2D and 3D~\cite{PhysRevB.98.174421,Shao2017b,Huang2017,qin2017amplitude,GYSun2018}. Therefore the techniques for investigating the dynamical properties of the QED$_3$-GN transitions are available. 

Aware of the high interests and great difficulty in studying the U(1) deconfined to confined transition, and with the help of the state-of-art QMC methodology and SAC machinery, in this work, we nevertheless take the first step to investigate the dynamical signature of the QED$_3$-GN transition at a large but finite fermion flavor of $N_f = 8$. As a function of the strength of the gauge fluctuations, the deconfined phase (the algebraic quantum spin liquid) and the confined phase (VBS) are investigated in detail, and the dynamical sigature of their transition in the form of the spin and dimer spectra with continua are discovered. The physical meaning of such continua inside the U(1) deconfined phase and at the transition are addressed as well. These results set the stage for the further investigations of the smaller and physically more relevant $N_f$ and can be used to guide the experimental detection in inelastic neutron scattering and nuclear magnetic resonance for condensed matter materials which could host fractionalized excitations coupled with emergent gauge structures~\cite{Feng2017,YuanWei2019}. For example, the observation of the conserved current correlations in the spin and dimer spectra~\cite{PhysRevLett.122.175701,JYLee2019,Huang2019} would be the decisive evidence for the deconfinement and emergent gauge fields.

With these thoughts in mind, we organize the rest of the paper as follows. In Sec.~\ref{sec:ii}, a quantum rotor model that describes the setting of compact QED$_3$ coupled to fermions is introduced. The QMC and SAC methods employed to solve the model are also explained in a concise manner. The analysis of the theoretical interpretation of the continua and associated symmetry properties of the U(1) deconfined phase are given in Sec.~\ref{sec:iii}. In Sec.~\ref{sec:iv}, QMC numerical results including the net gauge flux and most importantly, the spin and dimer spectra are presented, with the physical meaning of the continua therein thoroughly discussed, which serve as the dynamical signature of the U(1) phase and U(1)-to-VBS QED$_3$-GN transition.  Conclusion and outlook are presented in Sec.~\ref{sec:v}.

\section{Model and Method}
\label{sec:ii}
In this work we study a 2D quantum rotor model coupled to fermions considered in Ref.~\cite{Xiao2018Monte}, whose Hamiltonian can be written as
\begin{eqnarray}
\begin{split}
H=&\frac12 J N_f\sum_{\langle i,j\rangle}\frac14\hat L_{ij}^2-t\sum_{\langle i,j\rangle\alpha}\left(c_{i\alpha}^\dagger e^{\ii\hat\phi_{ij}}c_{j\alpha}+\text{h.c.}\right)\\
&+\frac12KN_f\sum_{\Box}\cos\left(\text{curl}\hat\phi\right),
\label{eq:eq1}
\end{split}
\end{eqnarray}
where $c_{i\alpha}$($c_{i\alpha}^\dagger$) is the annihilation (creation) operator for a fermion with fermion flavor $\alpha$. The $\alpha$ runs from 1 to $N_f$ and here we focus on the case of $N_f=8$. As shown in Fig.~\ref{fig:fig1} (b), the nearest hopping of fermions is associated with a phase $\phi_{ij}$, this phase inserts magnetic flux through each plaquette. The flux term with $K > 0$ favors
$\pi$-flux in each elementary plaquette $\Box$. Following the convention in Ref.~\cite{Xiao2018Monte}, we fixed $t=1$ and $K=1$ and scan the $J$-axis.
$\hat L_{ij}$ is the canonical angular momentum operator and it satisfies the commutation relation of $[\hat L_{ij},e^{\pm i\hat\phi_{ij}}]=\pm e^{\pm i\hat \phi_{ij}}$, and $J$ is the strength of the gauge field fluctuations. The overall phase diagram of Eq.~\eqref{eq:eq1} is obtained in the previous QMC work~\cite{Xiao2018Monte} and is adapted here in Fig.~\ref{fig:fig1} (a). 
	
In the quantum Monte Carlo simulation in Ref.\cite{Xiao2018Monte}, the
quantum critical points can be extracted by means of correlation ratio, which is defined as $
r= 1- \frac{\chi(\mathbf{X}+\delta{\mathbf{q}})}{\chi(\mathbf{X})}$,
where $\mathbf{X}$ is the order wavevector for AFM ($\mathbf{X}=(\pi,\pi)$) or VBS ($\mathbf{X}=(\pi,0)$) on the square lattice and $\delta(\mathbf{q})=(\frac{2\pi}{L},0)$ is the smallest momentum away from $\mathbf{X}$. The $\chi$ is the correlation function of the corresponding order that one probes, for example, the AFM order is determined by the $\chi$ of spin-spin correlation function $\chi_{S}(\vec{k})=\frac{1}{L^{4}}\sum_{ij}\sum_{\alpha\beta}\langle
S_{\beta}^{\alpha}(i)S_{\alpha}^{\beta}(j)\rangle
e^{i\vec{k}\cdot(\vec{r}_{i}-\vec{r}_{j})}$ where the spin operator
$S_{\beta}^{\alpha}(i)=c_{i\alpha}^{\dagger}c_{i\beta}-\frac{1}{N_{f}}\delta_{\alpha\beta}\sum_{\gamma}c_{i\gamma}^{\dagger}c_{i\gamma}$, and the VBS order is determined by the $\chi$ of dimer-dimer correlation function $\chi_{D}(\vec{k})=\frac{1}{L^{4}}\sum_{ij}\left(\langle
D_{i}D_{j}\rangle-\langle D_{i}\rangle\langle
D_{j}\rangle\right)e^{i\vec{k}\cdot(\vec{r}_{i}-\vec{r}_{j})}$ with the dimer operator
$D_{i}=\sum_{\alpha\beta}S_{\beta}^{\alpha}(i)S_{\alpha}^{\beta}(i+\hat{x})$
is defined as dimer along the nearest-neighbor bond in $\hat{x}$
direction.
 
By monitoring the corresponding correlation ratios, Ref.~\cite{Xiao2018Monte} gives that for $N_f=2$, the transition of U1D-to-AFM is at $J=1.6(2)$; for $N_f=4$, the transition of U1D-to-VBS is at $J=1.2(3)$, and the transition of VBS-to-AFM is at $J=17(4)$; for $N_f=6$, the transition of U1D-to-VBS is at $J=1.9(3)$; for $N_f=8$, the transition of U1D-to-VBS is at $J=2.5(1)$. The illustration of the model in Eq.~\eqref{eq:eq1} is shown in Fig.~\ref{fig:fig1} (b).

This compact U(1) lattice gauge theory coupled to fermionic matter at (2+1)D is now accessible with large-scale QMC simulations.  The Hamiltonian in Eq.~\eqref{eq:eq1} can be formulated in
a coherent-state path integral. To simulate the above model with determinantal QMC method, we start with the partition function as detailed in Ref.~\cite{Xiao2018Monte,XYXu2019},
\begin{equation}
Z=\Tr e^{-\beta H}=\sum_{[\phi]}\omega_B[\phi]\omega_F[\phi],
\end{equation}
where the configuration space of $[\phi]$ is comprised of the (2+1)D gauge field. The bosonic  part  of the partition function is
\begin{equation}
\begin{split}
\omega_B[\phi]=\prod_{\tau=M}^1&e^{-\frac{4}{JN_f\Delta\tau}\sum_{\langle ij\rangle}\Big(1-\cos\big(\phi_{ij}(\tau+1)-\phi_{ij}(\tau)\big)\Big)}\\
&e^{-\frac{1}{2}K N_f \Delta\tau\cos\big(\sum_{\langle i,j \rangle \in \Box}\phi_{i,j}(\tau)\big)},
\end{split}
\end{equation}
and the fermionic part of the partition function is
\begin{equation}
\omega_F[\phi]=\left(\det\left[I+B^MB^{M-1}\cdots B^\tau\cdots B^2B^1\right]\right)^{N_f}.
\end{equation}
Here all the flavors of fermion are subject to the same gauge field configuration, so for every fermion flavor, the $B^{\tau}$ matrix in the fermionic weight $\omega_F[\phi]$ is given by
\begin{equation}
\begin{split}
B^\tau&=\prod_{\langle ij\rangle}\exp{\left(\Delta\tau\begin{bmatrix}
  0& & & &  \\
    &0& &e^{\ii\phi_{ij}(\tau)}& \\
  & &0 & & & \\
  &e^{-\ii\phi_{ij}(\tau)}& &0& \\
  & & & &0  \\
\end{bmatrix}\right)}\\
&=\prod_{\langle ij\rangle}\begin{bmatrix}
  1& & & &  \\
    &\cosh{\Delta\tau}&0&e^{\ii\phi_{ij}}\sinh{\Delta\tau}& \\
  &0 &1 &0 & & \\
    &e^{-\ii\phi_{ij}}\sinh{\Delta\tau}&0&\cosh{\Delta\tau}& \\
  & & & &1  \\
\end{bmatrix}.
\end{split}
\end{equation}
Since the gauge field $\phi_{i,j}(\tau)$ are continuous variables at the $(2+1)$D space time, and matrix elements in $B^{\tau}$ are complex numbers, it is very important to use an efficient strategy to update the gauge field $[\phi]$~\cite{Xiao2018Monte}. We update the U(1) gauge field on $l$-th imaginary-time slice at $ij$-th lattice bond from $\phi_{ij}$ to $\phi'_{ij}$. The ratio which determines whether we accept
the updating can be expressed as
\begin{equation}
r=\frac{\omega_B[\phi']}{\omega_B[\phi]} \cdot \frac{\omega_F[\phi']}{\omega_F[\phi]}.
\end{equation}
For the boson part, the ratio of the weight is
\begin{widetext}
\begin{eqnarray}
\begin{split}
\frac{\omega_B[\phi']}{\omega_B[\phi]}=&\frac{\exp\left({\frac{4}{JN_f\Delta\tau}\left(\cos{\left(\phi_{ij}\left(\tau+1\right)-\phi'_{ij}\left(\tau\right)\right)}+\cos\left(\phi_{ij}\left(\tau-1\right)-\phi'_{ij}\left(\tau\right)\right)\right)}\right)}{\exp\left({\frac{4}{JN_f\Delta\tau}\left(\cos{\left(\phi_{ij}\left(\tau+1\right)-\phi_{ij}\left(\tau\right)\right)}+\cos\left(\phi_{ij}\left(\tau-1\right)-\phi_{ij}\left(\tau\right)\right)\right)}\right)}\cdot\frac{\exp\left(-\frac12\Delta\tau K N_f\cos\left(\sum_{\langle ij\rangle \in\Box }\phi'_{i,j}(\tau)\right)\right)}{\exp\left(-\frac12\Delta\tau K N_f\cos\left(\sum_{\langle ij\rangle \in\Box }\phi_{i,j}(\tau)\right)\right)}\\
=&\exp\Bigg[-\frac{16}{JN_f\Delta\tau}\cdot\cos\left(\frac{\phi_{ij}(\tau+1)-\phi_{ij}(\tau-1)}{2}\right)\cdot\sin\left(\frac{\phi_{ij}(\tau+1)+\phi_{ij}(\tau-1)-\phi'_{ij}(\tau)-\phi_{ij}(\tau)}{2}\right)\cdot\\
&\sin\left(\frac{\phi_{ij}(\tau)-\phi'_{ij}(\tau)}{2}\right)-\frac{\Delta\tau K N_f}{2}\left(\cos\left(\sum_{\langle ij\rangle\in\Box}\phi'_{ij}(\tau)\right)-\cos\left(\sum_{\langle ij\rangle\in\Box}\phi_{ij}\left(\tau\right)\right)\right)\Bigg],
\end{split}
\end{eqnarray}
\end{widetext}
and for the fermionic part, the ratio of the weight is
\begin{eqnarray}
\begin{split}
\frac{\omega_F[\phi']}{\omega_F[\phi]}&=\frac{\det\left[I+B(\beta,\tau)(1+\Delta)B(\tau,0)\right]}{\det\left[I+B(\beta,\tau)B(\tau,0)\right]}\\
&=\det\{1+\Delta\left[1-G(\tau,\tau)\right]\}.
\end{split}
\end{eqnarray}
If the update is accepted, we also need update equal-time Green's function as
\begin{eqnarray}
G'(\tau,\tau)=G(\tau,\tau)\left[1+\Delta(1-G(\tau,\tau))\right]^{-1},
\end{eqnarray}
with the $2\times2$ matrix of $\Delta$
\begin{equation}
\Delta=\begin{bmatrix}
  \Delta_{ii} & \Delta_{ij} \\
  \Delta_{ji} & \Delta_{jj} \\
\end{bmatrix},
\end{equation}
where
\begin{equation}
\begin{split}
&\Delta_{ii}= 1-e^{-\ii\left(\phi_{ij}-\phi'_{ij}\right)}\sinh^2\Delta\tau,\\
&\Delta_{jj}=1-e^{\ii\left(\phi_{ij}-\phi'_{ij}\right)}\sinh^2\Delta\tau,\\
&\Delta_{ij}= \left(-e^{\ii\phi_{ij}}+e^{\ii\phi'_{ij}}\right)\sinh\Delta\tau\cosh\Delta\tau,\\
&\Delta_{ji}= \left(-e^{-\ii\phi_{ij}}+e^{-\ii\phi'_{ij}}\right)\sinh\Delta\tau\cosh\Delta\tau.
\end{split}
\end{equation}
The more detail of DQMC method used in this work can be found in Ref.~\cite{Xiao2018Monte}. It is with such QMC methodology that the ground state phase diagram as a function of fermion flavor $N_f$ and the
strength of gauge fluctuations $J$ is mapped out, as shown in Fig.~\ref{fig:fig1} (a).

In this paper, we focus on the large fermion flavor case ($N_f=8$) by means of stochastic analytic continuation (SAC)
of imaginary-time correlation functions obtained from DQMC, where the deconfine-confine phase transition is investigated in detail~\cite{Xiao2018Monte}, the  $1/L$ extrapolation of the correlation ratio crossings estimates U1D-to-VBS transition point
at $J_c = 2.5(1)$ for $N_f = 8$.

The time displaced correlated function (defined as $G_{i,j}(\tau)=\langle\hat O_i(\tau)\hat O_j(0)\rangle$ and $G(\mathbf{q},\tau)=\frac{1}{N^2}\sum_{i,j}e^{\ii\mathbf{q}\cdot\mathbf{r}_{i,j}}G_{i,j}(\tau)$) of an operator $\hat O$ for a set of imaginary times $\tau_i \ (i=0,1,\cdots,N_\tau)$ with statistical errors can be obtained from DQMC simulations. By SAC method~\cite{PhysRevE.94.063308,Shao2017b,Huang2017,qin2017amplitude}, the corresponding
real-frequency spectral function $A(\omega)$ can be obtained from them according to the relationship of $G(\tau)=\int_{-\infty}^\infty d\omega A(\omega)K(\tau,\omega)$, where the kernel $K(\tau,\omega)$ depends on the type of the spectral function, i.e., fermionic or bosonic, finite or zero temperature. The spectra at positive and negative frequencies obey the relation of $A(-\omega)=e^{-\beta\omega}A(\omega)$ and we are restricted at the positive frequencies and the kernel can then be written as $K(\tau,\omega)=\frac1\pi(e^{-\tau\omega}+e^{-(\beta-\tau)\omega})$.
In order to work with a spectral function that is itself normalized to unity on the positive frequency axis, we modify the kernel and the spectral function and arrive at the transformation between the imaginary time Green's function $G(\mathbf{q},\tau)$ and real-frequency spectral function $B(\mathbf{q},\omega)$
\begin{eqnarray}
\label{eq:eq7}
G(\mathbf{q},\tau)=\int_0^\infty \frac{d\omega}{\pi}\frac{e^{-\tau\omega}+e^{-(\beta-\tau)\omega}}{1+e^{-\beta\omega}}B(\mathbf{q},\omega)
\end{eqnarray}
where $B(\omega)=A(\omega)(1+e^{-\beta\omega})$.

In the practical calculation, we parametrize the $B(\omega)$ with a large number of equal-amplitude
$\delta$-functions sampled at locations in a frequency continuum as $B(\omega)=\sum_{i=0}^{N_\omega-1}a_i\delta(\omega-\omega_i)$. Then the relationship between Green's function obtained from Eq.~\eqref{eq:eq7} and from DQMC can be described by the goodness of fit $\chi^2$, i.e. $\chi^2=\sum_{i=1}^{N_\tau}\sum_{j=1}^{N_\tau} (G_i-\bar G_i) C_{ij}^{-1}(G_j-\bar G_j)$, where $\bar G_i$ is the average of DQMC measurement and $C_{ij}$
is covariance matrix $C_{ij}=\frac{1}{N_B(N_B-1)}\sum_{b=1}^{N_B}(G_i^b-\bar G_i)(G_j^b-\bar G_j)$. Here $N_B$ is the number of bins in the measurement of DQMC.
Then we update the series of $\delta$-functions in a Metropolis process, from $(a_i,\omega_i)$ to $(a'_i,\omega'_i)$, to get a more probable configuration of $B(\omega)$. The weight for a given spectrum follows the Boltzmann distribution $P(B)\propto \exp(-\chi^2/2\Theta)$, with $\Theta$ a fictitious
temperature chosen in an optimal way so as to give a statistically sound mean $\chi^2$ value, while still staying in the
regime of significant fluctuations of the sampled spectra
so that a smooth averaged spectral function is obtained. The resulting spectra will be collected as an ensemble average of the Metropolis process within the configurational space of $\{a_i, \omega_i\}$, as explained in Refs.~\cite{PhysRevB.98.174421,PhysRevE.94.063308,Shao2017b,Huang2017,qin2017amplitude,GYSun2018}.

\section{Field Theory Analysis of U(1) Deconfined Phase}
\label{sec:iii}
\subsection{$\pi$-Flux State Mean-Field Theory}
Before presenting our numerical result, we first provide a mean-field study of the spin and dimer excitation spectra in the U1D phase ignoring the $\U(1)$ gauge fluctuation (or considering the $J\to 0$ limit of \eqnref{eq:eq1} model). The mean-field treatment will be asymptotically exact in the large $N_f$ limit. At the mean-field saddle point, the fermions experiences a $\pi$-flux (per plaquette) background, described by the following Hamiltonian on the square lattice
\begin{equation}\label{eq:HMF}
H_\text{MF}=t\sum_i \ii(c_{i+\hat{x}}^\dagger c_i+(-)^x c_{i+\hat{y}}^\dagger c_i)+h.c.
\end{equation}
where $c_i,c_i^\dagger$ are the creation and annihilation operators for the fermions on site $i$ with $N_f=8$ internal flavors. The $N_f$ lattice fermions will give rise to $2N_f$ Dirac fermions at low energy following the fermion doubling theorem. To see this, we transform the Hamiltonian to the momentum space,
\begin{equation}\label{eq:HMF in k}
H_\text{MF}=-2t\sum_\vect{k} c_\vect{k}^\dagger (\sin k_x \sigma^{10}+\sin k_y \sigma^{31})\otimes\mathds{1}_{8\times 8} c_\vect{k},
\end{equation}
where we have chosen the four-site unit cell (sublattices are arranged surrounding a plaquette) such that $c_\vect{k} = (c_{\vect{k}A},c_{\vect{k}B},c_{\vect{k}C},c_{\vect{k}D})^\intercal$ and each component $c_{\vect{k}a}$ further contains $N_f = 8$ flavors. Here $\sigma^{ij\cdots}\equiv \sigma^i \otimes \sigma^j \otimes\cdots$ denotes the tensor product of Pauli matrices and $\sigma^0$ stands for $2\times 2$ identity matrix. $\mathds{1}_{8 \times 8}$ is identity matrix with dimension $8\times 8$, we will use $\mathds{1}$ for short. The fermion dispersion is given by
$\epsilon_\vect{k}=\pm 2 t(\sin^2 k_x + \sin^2 k_y)^{1/2}$, which is gapless at the momentum $\vect{k}=(0,0)$. Expand around the Dirac point and rescale the theory to eliminate $t$, the low-energy continuum model can be written in terms of the Lagrangian density as,
\begin{equation}\label{equ:lagrangian}
    \mathcal{L} =\bar{c} (\ii \gamma^\mu \partial_\mu)c,
\end{equation}
where $\gamma^\mu = (\ii \sigma^{21}\mathds{1}, -\sigma^{31}\mathds{1}, \sigma^{10}\mathds{1})$ and $\bar c = c^\dagger \gamma^0$. Note that the gamma matrices are of dimensions $4N_f\times 4N_f$. Given that the minimal dimension of gamma matrices for (2+1)D Dirac fermion is $2\times 2$, the above flavor counting confirms that the $\pi$-flux model contains $2N_f$ Dirac fermions at low energy.

\subsection{Spin and Dimer Excitation Spectrum}
\label{sec:iiib}
Given $N_f=8$ fermions on each site, our model has the $\SU(8)$ flavor symmetry on the lattice level. The $\SU(8)$ spin operators are defined as
\begin{equation}\label{eq:SU(8)spin}
S_i^a=\frac{1}{2}c_i^\dagger T^a c_i,
\end{equation}
where $T^a$ (with $a=1,\cdots, 63$) are the $N_f^2-1=63$ generators of $\SU(8)$. The dimmer operators along $x$ and $y$ directions are defined as $D_i^x=\sum_{a}S_i^aS_{i+\hat{x}}^a$ and $D_i^y=\sum_{a}S_i^aS_{i+\hat{y}}^a$ respectively. One may expect to expand them to four-fermion operators by inserting \eqnref{eq:SU(8)spin}. However, the following fermion bilinear operators
\begin{equation}\label{eq:dimer}
\begin{split}
D_i^x &= -(c_{i+\hat{x}}^\dagger e^{\ii\phi_{i+\hat{x},i}} c_i +h.c.) /2\\&=(\ii c_{i+\hat{x}}^\dagger c_i +h.c.) /2,\\
D_i^y &=-(c_{i+\hat{y}}^\dagger e^{\ii\phi_{i+\hat{y},i}} c_i +h.c.) /2\\&=(-)^x (\ii c_{i+\hat{y}}^\dagger c_i +h.c.) /2,
\end{split}
\end{equation}
are gauge invariant (where we have replaced the dynamic gauge connection $\phi_{ij}$ by its mean field value specified in \eqnref{eq:HMF}) and symmetry-wise equivalent to the dimmer operators. In the large $N_f$ limit, the fermion bilinear operators are generally more relevant at low energy. So under the renormalization group flow, the dimmer operators should be represented by the fermion bilinear operator in \eqnref{eq:dimer}.

The Fourier transform for generic operator is defined via $\mathcal{O}_\vect{q} = \sum_i \mathcal{O}_i e^{-\ii \vect{q}\cdot \vect{r}_i}$. For fermion bilinear operators, they take the general form of
\begin{equation}
\mathcal{O}_\vect{q}=\sum_\vect{k} c_\vect{k}^\dagger v_\vect{q} c_{\vect{k}+\vect{q}},
\end{equation}
where $v_\vect{q}$ is the vertex function (matrix) that depends on the momentum transfer $\vect{q}$. When $\vect{q}$ goes beyond the fermion Brillouin zone, we apply the following rules to map $c_{\vect{k}+\vect{q}}$ back: $c_{\vect{k}+(\pi,0)}=\sigma^{30}\mathds{1}c_\vect{k}$, $c_{\vect{k}+(0,\pi)}=\sigma^{03}\mathds{1}c_\vect{k}$, $c_{\vect{k}+(\pi,\pi)}=\sigma^{33}\mathds{1}c_\vect{k}$. Applying to the spin and dimer operators, we explicitly have
\begin{equation}\label{eq:operator in q}
\begin{split}
    S_\vect{q}^a &= \frac{1}{2}\sum_\vect{k} c_\vect{k}^\dagger \sigma^{00} T^a c_{\vect{k}+\vect{q}},\\
    D_\vect{q}^x &= \sum_\vect{k} c_\vect{k}^\dagger (\sin (k_x+\frac{q_x}{2}) e^{-\ii q_x/2}\sigma^{10}\mathds{1})c_{\vect{k}+\vect{q}},\\
    D_\vect{q}^y &= \sum_\vect{k} c_\vect{k}^\dagger (\sin (k_y+\frac{q_y}{2}) e^{-\ii q_y/2}\sigma^{31}\mathds{1})c_{\vect{k}+\vect{q}}.
\end{split}
\end{equation}
With these, we can evaluate the correlation function for spin or dimer operators
\begin{equation}
    \chi_\mathcal{O}(\vect{q},\omega)=\int dt \  e^{\ii \omega t} \langle \mathcal{O}_{-\vect{q}}(t)\mathcal{O}_\vect{q}(0) \rangle,
\end{equation}
from which the spectral function $A_\mathcal{O}(\vect{q},\omega)=-2\Im\chi_\mathcal{O}(\vect{q},\omega+\ii 0_+)$ can be obtained. In the zero temperature limit, the spectral function is given by
\begin{equation}
\begin{split}
A_\mathcal{O}(\vect{q},\omega)&=\sum_{m,n,\vect{k}}\bra{m,\vect{k}}v_\vect{-q}\ket{n,\vect{k}+\vect{q}}\bra{n,\vect{k}+\vect{q}}v_\vect{q}\ket{m,\vect{k}}\\
&\delta(\omega+\xi_{m,\vect{k}}-\xi_{n,\vect{k}+\vect{q}})(\Theta(\xi_{m,\vect{k}})-\Theta(\xi_{n,\vect{k}+\vect{q}}))
\end{split}
\end{equation}
where $\ket{n,\vect{k}}$ is the $n$-th eigenstate of the single-particle Hamiltonian \eqnref{eq:HMF in k} with momentum $\vect{k}$ and $\xi_{n,\vect{k}}$ is the corresponding eigen energy, and $\Theta$ denotes the step function. Given the $v_\vect{q}$ in Eq.~\eqref{eq:operator in q}, the above calculation will provide us the understanding of the overall shape of the spectral function for both spin and dimer correlations in the $J\to0$ limit of the lattice QED model in \eqnref{eq:eq1}, where $\U(1)$ gauge fluctuation is supressed. We will demonstrate the spectra in Fig.~\ref{fig:fig3} and Fig.~\ref{fig:fig5} in Sec.\ref{sec:iv}, and compare them with our QMC result involving gauge fluctuations. We find that the low-energy spectral features match nicely on the qualitative level between the free fermion and the QMC results (although the scaling dimensions will be altered by gauge fluctuations).

\subsection{Emergent Symmetry and Conserved Currents}
\label{sec:iiic}
The mean-field understanding of the excitation spectrum helps us to identify signatures of emergent symmetry in the U1D phase. Let us restore the $\U(1)$ gauge fluctuation in the following discussion, and consider the compact QED theory with $16$ Dirac fermions in $(2+1)$D spacetime. The Lagrangian in \eqnref{equ:lagrangian} becomes
\begin{equation}\label{eq:QED}
\mathcal{L}=\bar{c}\gamma^\mu(\ii\partial_\mu-a_\mu)c+g(\mathcal{M}+\mathcal{M}^\dagger)+\cdots,
\end{equation}
where $\mathcal{M}$ and $\mathcal{M}^\dagger$ are the annihilation and creation operators of the $2\pi$ flux of the $\U(1)$ gauge field, also known as the monopole operator (event) in the spacetime.\cite{XYSong2018,XYSong2018b} Such monople terms are generally allowed if not forbidden by the physical symmetry. Here the physical symmetry includes the spin $\SU(8)$ symmetry and the four-fold rotation symmetry $\mathbb{Z}_4$ of the square lattice, which act on the fermion field $c$ as
\begin{equation}
\SU(8):c\to e^{\ii \theta_a T^a}c,\quad\mathbb{Z}_4:c\to e^{\ii\frac{\pi}{4}\tau^3}c,
\end{equation}
where $\tau^3=\sigma^{12}\mathds{1}$ generates the rotation between $x$-VBS and $y$-VBS. Based on operator-state correspondence, the $\U(1)$ gauge monopole operator with charge 1 can be effectively mapped to the state on $S^2\times \mathbb{R}$ with 1 unit of background magnetic flux through $S^2$, and the states contain fermion zero mode guaranteed by Atiyah-Singer index theorem \cite{Borokhov2002Topological}. Therefore, when a $\U(1)$ monopole operator is inserted, each Dirac cone will contribute a zero mode, so there are totally 16 zero modes. Different ways of filling these zero modes leads to different monopole states that are degenerated in energy. A gauge neural monopole must have these fermion zero modes half-filled (i.e.~filling 8 fermions on 16 zero modes), which results in $C_{16}^{8}=12870$ different monopole states. Among them, only $9$ states preserves the spin $\SU(8)$ symmetry. They can be labeled by the following quantum number
\begin{equation}
m=\frac{1}{2}c^{\dagger}\tau^3 c=0,\pm1,\pm2,\pm3,\pm4,
\end{equation}
which corresponds to the monopole angular momentum because $\tau^3/2$ is the generator of  the $\mathbb{Z}_4$ lattice rotation symmetry. For example, the $m=4$ state is created by $\mathcal{M}^\dagger\sim\prod_{\alpha=1}^8 \frac{1+\tau^3}{2}c_\alpha^\dagger$, which fills all the 8 fermions on the monopole modes of the same valley (of $\tau^3=+1$). Each fermion occupies a distinct $\SU(8)$ spin flavor, such that the monopole state is $\SU(8)$ symmetric. Further imposing the $\mathbb{Z}_4$ rotation symmetry to the monopole, its angular momentum must satisfy $m=0\mod 4$, which further singles out 3 monopole states labeled by $m=0,\pm4$. These states span the Hilbert space of a single monopole that preserve all the $\SU(8)\times\mathbb{Z}_4$ physical symmetry, so their corresponding monopole operators are generally allowed to appear in the Lagrangian of the QED theory \eqnref{eq:QED}. 

Depending on the relevance of the single monopole term $g$ at the large-$N_f$ fixed point, the lattice QED model in \eqnref{eq:eq1} can have different emergent symmetry in the U1D phase. The scaling dimension $\Delta_\mathcal{M}$ of the single monopole operator has been calculated in the large-$N_f$ limit in Ref.\,\onlinecite{Borokhov2002Topological,Dyer2013Monopole,Pufu2013Monopoles}, which reads $\Delta_\mathcal{M}=0.265 \times(2N_f)-0.0383+\mathcal{O}(1/N_f)$ \footnote{Note that the notion of $N_f$ in the lattice model is differed from that in the field theory by a factor of 2}. With $N_f=8$, $\Delta_\mathcal{M}=4.2>3$, so the single monopole operator is irrelevant to the leading orders in $1/N_f$, nevertheless the conclusion may still change at higher orders of $1/N_f$. But if we accept that the single monopole term is \emph{irrelevant}, the theory will flow to the QED$_3$ fixed point, where the emergent symmetry is the full $\SU(16)/\mathbb{Z}_{16}=\PSU(16)$ flavor symmetry of the $16$ Dirac fermions, where the $\mathbb{Z}_{16}$ center of the $\SU(16)$ group should be quotient out because this subgroup is shared with the $\U(1)$ gauge group. The $\SU(16)$ generators can be enumerated as $\{\vect{\tau},T^a,\vect{\tau}T^a\}$. Here $\vect{\tau}=(\sigma^{01}\mathds{1},\sigma^{13}\mathds{1},\sigma^{12}\mathds{1})$ are the generators of valley rotations. These generators are found by requiring them to commute with $\gamma^\mu$, such that the Lagrangian in \eqnref{eq:QED} remains invariant under the fermion flavor rotation. Using the $\SU(16)$ generators, one can define the $\SU(16)$ currents (labeled by $i=1,2,3$ and $a=1,\cdots,63$)
\begin{equation}\label{eq:current}
j_{i0}^\mu=\bar{c}\gamma^\mu\tau^i c,j_{0a}^\mu=\bar{c}\gamma^\mu T^a c,j_{ia}^\mu=\bar{c}\gamma^\mu\tau^i T^a c.
\end{equation}
There are $255$ current operators in total (each current further contains 3 spacetime components labeled by $\mu=0,1,2$). All these currents are emergent conserved currents at low energy.

However, although unlikely, if the single monopole operator turns out to be \emph{relevant} and if we assume the theory flows to another non-trivial conformal fixed point (when $N_f$ is within the conformal window), the emergent symmetry can be lowered by the non-vanishing monopole term. The single monopole term $g$ will break the emergent symmetry from $\PSU(16)$ to
\begin{equation}
\frac{\SU(8)}{\mathbb{Z}_8}\times\frac{\SU(8)}{\mathbb{Z}_8}\times\frac{\mathbb{Z}_8}{\mathbb{Z}_2}=\PSU(8)\times\PSU(8)\times\mathbb{Z}_4.
\end{equation}
The above symmetry group is most easily seen for the $m=4$ monopole: the two $\SU(8)$ acts on the spin flavors in the two valleys ($\tau^3=\pm1$) respectively, $\mathbb{Z}_8$ is the opposite 8-fold rotation of fermion phases in opposite valleys, and all the center subgroups must be quotient out as they are shared between the $\U(1)$ gauge group. More careful symmetry analysis for the other monopoles of different $m$ shows that the above symmetry is indeed the largest possible residual symmetry of a single monopole operator. In this case, the emergent conserved currents are reduced to \begin{equation}
j_{0a}^\mu=\bar{c}\gamma^\mu T^a c,j_{3a}^\mu=\bar{c}\gamma^\mu\tau^3T^a c.
\end{equation}
In this case, there are $126$ emergent conserved currents in total.
We summarize the above analysis in \tabref{tab:emergent}.

\begin{table}[htp]
\caption{Emergent symmetry and conserved currents for the $N_f=8$ lattice QED model.}
\begin{center}
\begin{tabular}{c|cc}
Monopole operator & irrelevant & relevant \\
\hline
Emergent symmetry & $\PSU(16)$ & $\PSU(8)\times\PSU(8)\times\mathbb{Z}_4$\\
Conserved currents & $j_{i0}^\mu,j_{0a}^\mu,j_{ia}^\mu$ & $j_{0a}^\mu,j_{3a}^\mu$\\
Number of currents & 255 & 126
\end{tabular}
\end{center}
\label{tab:emergent}
\end{table}

Our analysis shows that the relevance of the $\U(1)$ gauge monopole crucially affects the emergent symmetry and the emergent conserved currents that can be probed at low-energy. Identifying these current fluctuations in the spin and dimer excitation spectra will be the first step towards pinning down the emergent symmetry and studying the monopole effects in the lattice QED model. The analysis can be carried out on the mean-field level. According to \eqnref{eq:operator in q} and $\bar{c}=c^\dagger \gamma^0$, $\gamma^\mu=(\ii\sigma^{21}\mathds{1},-\sigma^{31}\mathds{1},\sigma^{10}\mathds{1})$, $\tau^i=(\sigma^{01}\mathds{1},\sigma^{13}\mathds{1},\sigma^{12}\mathds{1})$, we can identify the following spin and dimer operators to current operators
\begin{equation}
\begin{split}
S_{(0,0)}^a&\sim c^\dagger \sigma^{00}T^a c=-\bar{c}\gamma^0T^ac=-j_{0a}^0,\\
S_{(\pi,0)}^a&\sim c^\dagger \sigma^{30}T^a c=\bar{c}\gamma^2\tau^1T^ac=j_{1a}^2,\\
S_{(0,\pi)}^a&\sim c^\dagger \sigma^{03}T^a c=\bar{c}\gamma^1\tau^2T^ac=j_{2a}^1,\\
D_{(\pi,\pi)}^x&\sim -c^\dagger \sigma^{23}\mathds{1} c=\bar{c}\gamma^2\tau^3 c=j_{30}^2,\\
D_{(\pi,\pi)}^y&\sim -c^\dagger \sigma^{02}\mathds{1} c=-\bar{c}\gamma^1\tau^3 c=-j_{30}^1.\\
\end{split}
\end{equation}
They are summarized in \tabref{tab:operators}. Among them, $j_{0a}^0$ is the conserved current of the physical spin $\SU(8)$ symmetry, and the remaining currents $j_{1a}^2,j_{2a}^1,j_{30}^2,j_{30}^1$ are all emergent conserved current of $\PSU(16)$ but not of $\PSU(8)\times\PSU(8)$, see \tabref{tab:emergent}. By measuring the scaling dimension of these currents from the spin and dimer correlation functions, one can decide if they are conserved or not to further determine the emergent symmetry and the relevance of the single monopole operator~\cite{PhysRevLett.122.175701}. At current stage, our numerical resolution is not sufficient to fully resolve the scaling dimension of these current fluctuations, nevertheless we will first map out the overall shape of the excitation spectra and identify these low-energy spectral features in this work and provide a road map for future study of the emergent conserved currents.

\begin{table}[h]
\caption{Identification of operators at high symmetry points.}
\centering
\begin{tabular}{ccccc}
 $\vect{q}$  & $(0,0)$ & $(\pi,0)$ & $(0,\pi)$ & $(\pi,\pi)$ \\
 \hline
 $S^a_\vect{q}$ & $-j_{0a}^0$ & $j_{1a}^2$ & $j_{2a}^1$ & N\'eel \\
 $D^x_\vect{q}$ & $\mathcal{L}-\mathcal{T}^{11}$  & $x$-VBS   & $-\ii \bar c \gamma^0 \tau^{2} \partial_x c$        & $j_{30}^2$ \\
 $D^y_\vect{q}$ & $\mathcal{L}-\mathcal{T}^{22}$  & $-\ii \bar c \gamma^0 \tau^{1} \partial_y c$  & $y$-VBS  & $-j_{30}^1$\\
\end{tabular}
    \label{tab:operators}
\end{table}

It is worth mentioning that the dimer fluctuation at $(0,0)$ momentum is gapless, but its spectral weight fades away much faster towards low energy as shown in Fig. \ref{fig:fig5} (a) and (b). This continuum corresponds to the energy-momentum tensor $T^{\mu\nu}$ which is the conserved current associated with the translation symmetry,
\begin{equation}\label{equ:d00-emtensor}
\mathcal{T}^{\mu\nu} = \frac{\partial \mathcal{L}}{\partial(\partial_\mu c)} \partial^\nu c -\delta^{\mu\nu} \mathcal{L}=\ii\bar{c}(\gamma^\mu\partial^\nu-\delta^{\mu\nu}\gamma^\lambda\partial_\lambda)c.
\end{equation}
Based on this definition, we can identify that
\begin{equation}
\begin{split}
D_{(0,0)}^x &\sim -\ii c^\dagger \sigma^{10}\mathds{1} \partial_x c=-\ii\bar{c}\gamma^{1}\partial_xc=\mathcal{L}-\mathcal{T}^{11},\\
D_{(0,0)}^y &\sim -\ii c^\dagger \sigma^{31}\mathds{1} \partial_y c=-\ii\bar{c}\gamma^{2}\partial_y c=\mathcal{L}-\mathcal{T}^{22}.
\end{split}
\end{equation}
The scaling dimensions of $\mathcal{L},\mathcal{T}^{\mu\nu}\sim\delta\mathcal{L}/\delta g_{\mu\nu}$ are 3, which follows from the fact that the action $S=\int\mathrm{d}^3 x\mathcal{L}$ and the metric $g_{\mu\nu}$ must be dimensionless. Because of the relatively high scaling dimension of the $D_{(0,0)}^x$ fluctuation, it is much more irrelevant under renormalization compared to the current and order parameter fluctuations. Therefore the low-energy spectral weight of $D_{(0,0)}^x$ is expected to be much weaker compare to other continua (e.g.~$D_{(\pi,0)}^x$, $D_{(\pi,\pi)}^x$) in the dimer excitation spectra in Fig.~\ref{fig:fig5} in Sec.~\ref{sec:iv}.

\section{Quantum Monte Carlo Results}
\label{sec:iv}
Here we present the QMC results, first begin with the definition of the physical observables.

\begin{figure}[htp]
\includegraphics[width =\columnwidth]{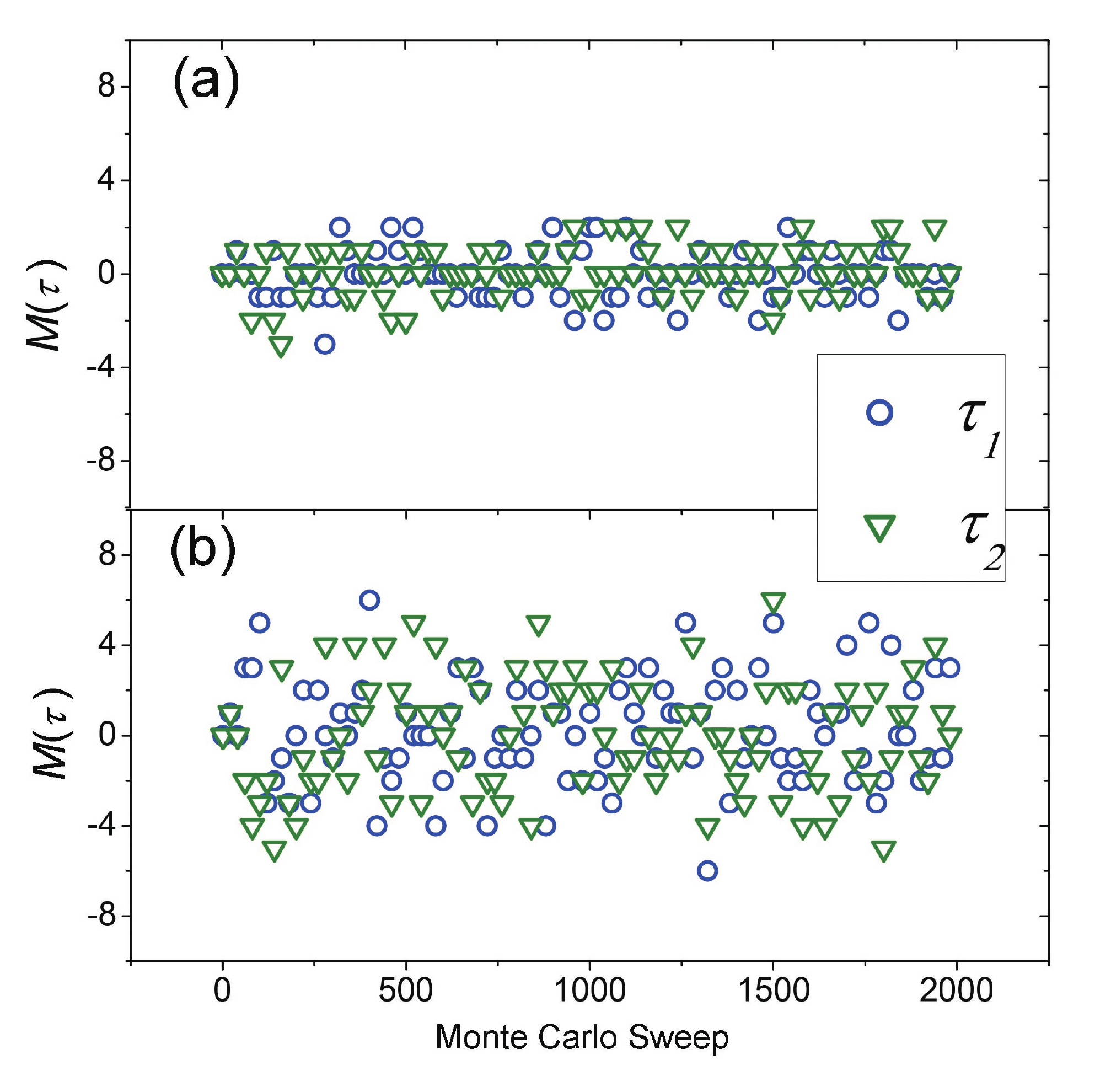}
\caption{Monte Carlo sweep serials of $M$, the fluctuations of the net flux at time slice $\tau_1$ and $\tau_2$ at with $L = 12$ (a) inside U1D phase at $J=1.80$ and (b) inside VBS phase at $J=3.50$. Here $\tau_1=\Delta\tau=0.05$ and $\tau_2=8\Delta\tau$. The flux sweep serials are plotted in the interval of 20 sweeps.} 
\label{fig:fig2}%
\end{figure}

\subsection{Physical observables}
 To understand the deconfine-confine phase transition, we focus on gauge-invariant dynamical structure factors obtained in QMC simulations, including the spin and dimer dynamical structure factor. They can be defined as the following forms\cite{Xiao2018Monte,Zhou2018Mott}
\begin{eqnarray}
S(\vect{q},\tau)=\frac{1}{N_s^2}\sum_{i,j}\sum_{\alpha,\beta}\left\langle S^\alpha_\beta(i,\tau)S^\beta_\alpha(j,0)\right\rangle e^{\ii\vect{q}\cdot(\mathbf{r}_i-\mathbf{r}_j)},
\end{eqnarray}
\begin{align}
D(\vect{q},\tau)&= \nonumber \\
&\frac{1}{N_s^2}\sum_{i,j}\left(\left\langle D_i(\tau)D_j(0)\right\rangle-\langle D_i(\tau)\rangle \langle D_j(0)\rangle\right) e^{\ii\vect{q}\cdot(\mathbf{r}_i-\mathbf{r}_j)}.
\end{align}
where the spin operator is $S_\beta^\alpha(i)=c_{i\alpha}^\dagger c_{i\beta}-\frac{1}{N_f}\delta_{\alpha\beta}\sum_{\gamma}c_{i\gamma}^\dagger c_{i\gamma}$
and the dimer operator is $D_i=\sum_{\alpha\beta}S_\beta^\alpha(i)S_\alpha^\beta(i+\hat x)$. The dimer operator is defined
along the nearest-neighbor bond in $\hat x$ direction.

\begin{figure*}[htp!]
\includegraphics[width =\textwidth]{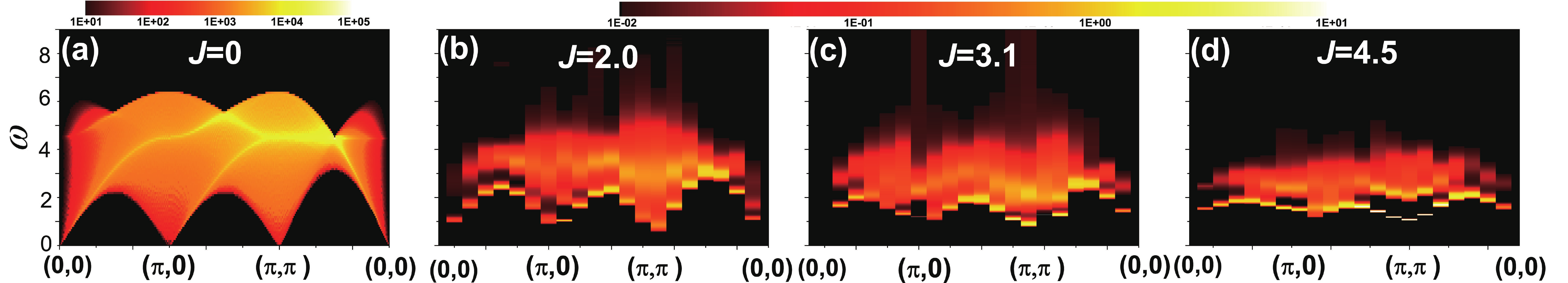}
\caption{(a) Spin spectra of non-interacting $\pi$-flux model and (b), (c), (d) the spectra obtained from QMC-SAC calculations through the phase transition of U1D-to-VBS with fermions flavors$N_f=8$, $L=14$ and $\beta=2L$.  (b) is inside the U1D phase with $J = 2.00 < J_c$, (c) is close to the critical point at $J=3.10$ and (d) is inside the VBS phase at $J = 4.50 > J_c$.}
\label{fig:fig3}%
\end{figure*}

As mentioned in Sec.~\ref{sec:ii}, the time displaced correlated functions $S(\vect{q},\tau)$ and $D(\vect{q},\tau)$ can be obtained in QMC for a set of imaginary times $\tau_i \ (i = 0, 1, \cdots, N_\tau )$ with statistical errors. From which, the SAC will be further applied to extract the real-frequency spectral functions $S(\vect{q},\omega)$ and $D(\vect{q},\omega)$.

\begin{figure*}[htp!]
\includegraphics[width=0.8\textwidth]{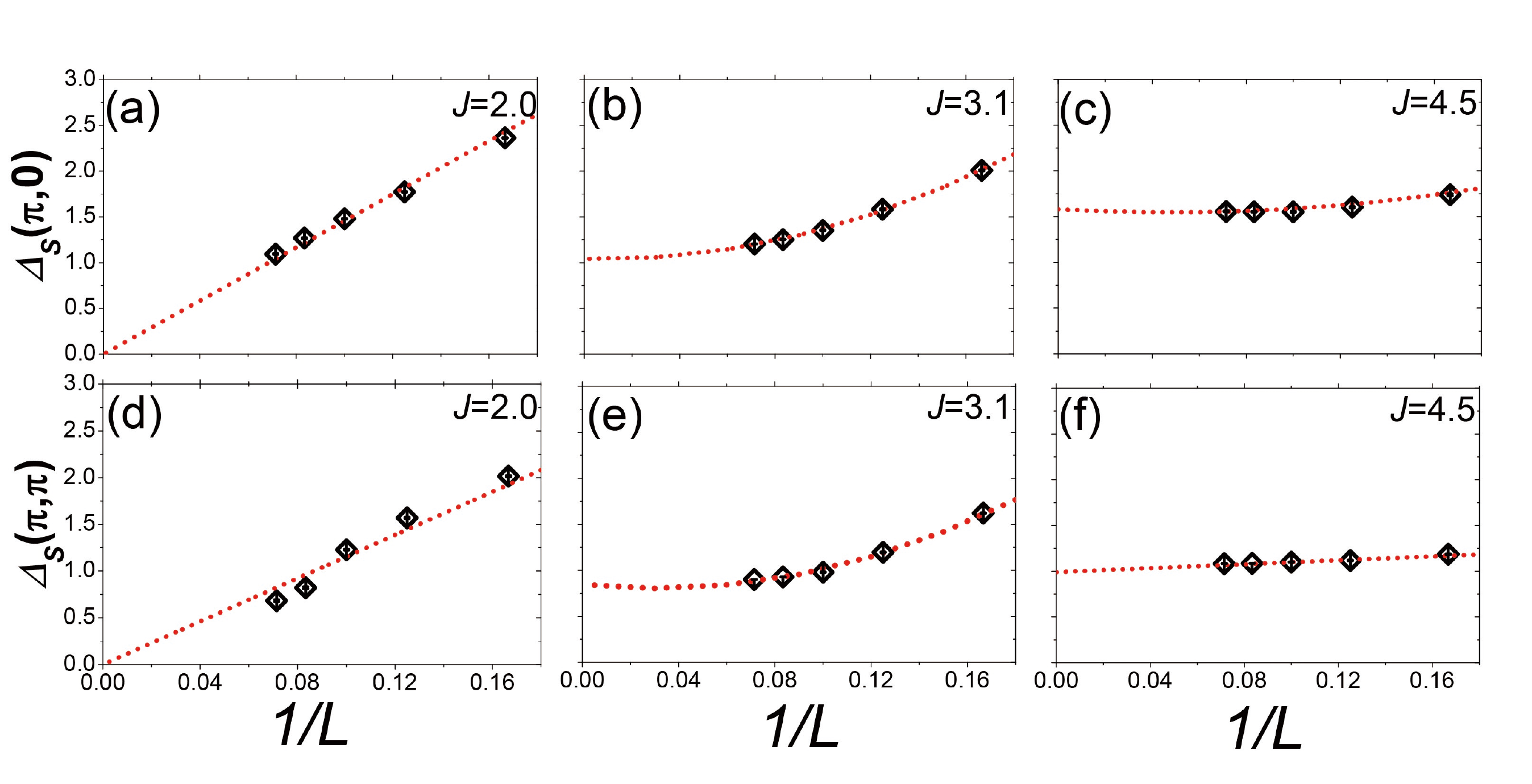}
\caption{The $1/L$ extrapolation of the spin gap at $(\pi,0)$ and $(\pi,\pi)$. (a) and (d) inside the U1D phase at $J = 2.00 < J_c$; (b) and (e) near the critical point at $J = 3.10 < J_c$; (c) and (f) inside the VBS phase at $J=4.50>J_c$.}
\label{fig:fig4}
\end{figure*}

Another quantity, that has distinctively different behaviors in U1D and VBS confined phases, is the net fluctuation of flux in each time slices with Monte Carlo steps~\cite{Xiao2018Monte,PhysRevLett.88.232001}. Flux in each plaquette can be written as $\sum_{b\in\Box}\phi_b=\Phi_{\Box}+2\pi m_\Box$ with $\Phi\in[0,2\pi)$ and $m_\Box$ an integer. The fluctuation of net flux
in one time slice $M(\tau)$ is defined as a sum of $m_\Box$ of each
plaquette at time slice $\tau$, $M(\tau)=\sum_{\Box}m_{\Box}(\tau)$. The evolution of $M(\tau)$ with Monte Carlo time series, both inside U1D and VBS phases at time slices  $\tau_1$  and $\tau_2$, are shown in Fig.~\ref{fig:fig2} (a) and (b), respectively. The parameters of calculation were given as $L=12$ and $\beta=2L$. Inside the U1D phase ($J=1.80<J_c$), as shown in Fig.~\ref{fig:fig2} (a) the net flux favors $\pi$ flux in each plaquette, and the net fluctuation $M(\tau)$ at each time slice seldom changes, $M(\tau_1)$ and $M(\tau_2)$ follow closely to each other and their value only take the integers $0$, $\pm 1$ and $\pm 2$; while in the VBS phase ($J=3.50 > J_c$), as shown in Fig.~\ref{fig:fig2} (b) the net fluctuations change almost randomly with more extended values, $0$, $\pm 1, \pm 2, \pm 3, \pm 4, \pm 5, \pm 6$ (in unit of $2\pi$), and large deviation between different time slices $\tau_1$ and $\tau_2$ can all be seen. These large fluctuations in the net flux indicate the proliferate of monopoles in the confined VBS phase.

\subsection{Spectra and excitation gaps}
In this part we present $S(\vect{q},\omega)$ and $D(\vect{q},\omega)$ inside both the U1D and VBS phases, these results are obtained from QMC-SAC simulations. We also show the corresponding spectra from the non-interacting $\pi$-flux model of Dirac fermions without gauge fluctuations, which are the spectra at the limit of $J=0$ derived in Sec.~\ref{sec:iiib}.

\subsubsection{Spin Spectra in UID and VBS phases}

Fig.~\ref{fig:fig3} shows the features of the spin spectra through the U1D-VBS transition, the results are shown along the high-symmetry-path of $(0,0)-(\pi, 0)-(\pi,\pi)-(0,0)$ . We present results for the non-interacting Dirac fermions corresponding to $J=0$ (Fig.~\ref{fig:fig3} (a)), inside the U1D phase with $J = 2.00 < J_c$ (Fig.~\ref{fig:fig3} (b)), close to the QED$_3$-GN critical point at $J=3.10$ (Fig.~\ref{fig:fig3} (c)) and inside the VBS confined phase at $J=4.50 > J_c$ (Fig.~\ref{fig:fig3} (d)).

The $\pi$-flux spinons, as discussed in Sec.~\ref{sec:iiib} with the U(1) gauge fluctuations suppressed, give rise to gapless spin spectra at momenta $(0,0)$, $(\pi,0)$ and $(\pi,\pi)$ in Fig.~\ref{fig:fig3} (a). The situation persists as one introduces the U(1) gauge fluctuations, for example at $J=2$ in Fig.~\ref{fig:fig3} (b). Of course on a finite size system $L=14$ for Fig.~\ref{fig:fig3} (b), the spectra look gapped due to finite size effect, we have performed the $1/L$ extrapolation of the spin gaps at $(\pi,0)$ and $(\pi,\pi)$ with the gaps directly obtained from fitting the imaginary time decay of $S(\vect{q},\tau)$ without SAC, the results are shown in Fig.~\ref{fig:fig4} (a) and (d), and it is clear that in the U1D phase, the spin excitations at $(\pi,0)$ and $(\pi,\pi)$ are gapless in the thermodynamic limit. As discussed in Sec.~\ref{sec:iiic}, the $(\pi,\pi)$ excitation corresponds to the spin $\SU(8)$ order parameter fluctuation, and the $(\pi,0)$ excitation corresponds to the current fluctuation whose charge operator generates the AFM-VBS rotation. If the emergent symmetry is $\PSU(16)$, the scaling dimension of the spin excitation at $(\pi,0)$ will be pinned at 2. However if the emergent symmetry is $\PSU(8)\times\PSU(8)\times\mathbb{Z}_4$, the scaling dimension will deviate from integer. More importantly, we also observed broad and prominent continuous spectra in Fig.~\ref{fig:fig3} (b), which reflects the expected deconfinement and fractionalization of spinons and their interactions mediated by the fluctuating U(1) gauge field. Similar $S(\vect{q},\omega)$, with gapless excitations at $(0,0)$, $(\pi,0)$ and $(\pi,\pi)$ and pronounced continua upto high energy, have also been seen at the deconfined quantum critical point with emergent O(4) symmetry~\cite{PhysRevB.98.174421,PhysRevLett.122.175701}.

As one moves towards the critical point (actually slightly above it in Fig.~\ref{fig:fig3} (c) with $J=3.1 > J_c=2.5$), broad and prominent continuous spectra can still be observed, signifying the effects of gauge fluctuations. And the $1/L$ extrapolation of the spin gap are shown in Fig.~\ref{fig:fig4} (b) and (e). It is clear now that once inside the VBS confined phase, the spin spectra are gapped due to the translational symmetry breaking of the VBS phase.
\begin{figure*}[htp!]
\includegraphics[width=\textwidth]{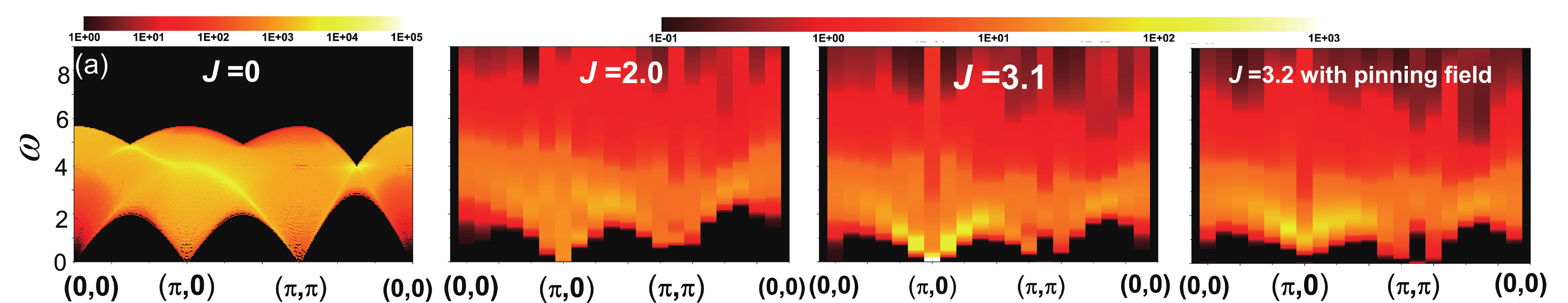}
\caption{(a) Dimer spectra from non-interacting $\pi$-flux model and (b), (c) and (d) are the spectra obtained from QMC-SAC calculations through the phase transition of U1D-to-VBS with fermions flavors $N_f = 8, L = 14$ and $ \beta= 2L$. (b) is in the U1D phase with $J = 2.00 < J_c$. (c) is close to the critical point at $J = 3.10$. (d) is in the VBS phase at $J = 3.20$ with pinning field.}
\label{fig:fig5}%
\end{figure*}
Deep inside the VBS phase, as shown in Fig.~\ref{fig:fig3} (d) with $J=4.5$, the spin spectra are fully gapped and the continua above it also become less extended in frequency domain. This is expected as well since here both the gauge fields and the fermions are interacting at the length scale shorter than that associated with the excitation gaps. Below the gap, the system is an insulator with fermions forming singlets along the $(\pi,0)$ or $(0,\pi)$ directions, i.e., translational symmetry breaking. The corresponding $1/L$ extrapolation of the gaps at $(\pi,0)$ and $(\pi,\pi)$ are shown in Fig.~\ref{fig:fig4} (c) and (f), respectively.

\subsubsection{Dimer Spectra in UID and VBS phases}
Fig.~\ref{fig:fig5} shows the dynamic dimer spectra
through the U1D-to-VBS transition.

The dimer spectra of the non-interacting $\pi$-flux Dirac fermions are given in Fig.~\ref{fig:fig5} (a). The spectra are gapless at momenta $(0,0)$, $(\pi,0)$ and $(\pi,\pi)$, similar to that of the spin in Fig.~\ref{fig:fig3} (a), but the spectral weights have different distribution. The calculation details of the non-interacting spin and dimer spectra are given in Sec.~\ref{sec:iiib}.

\begin{figure}[htp]
\includegraphics[width=\columnwidth]{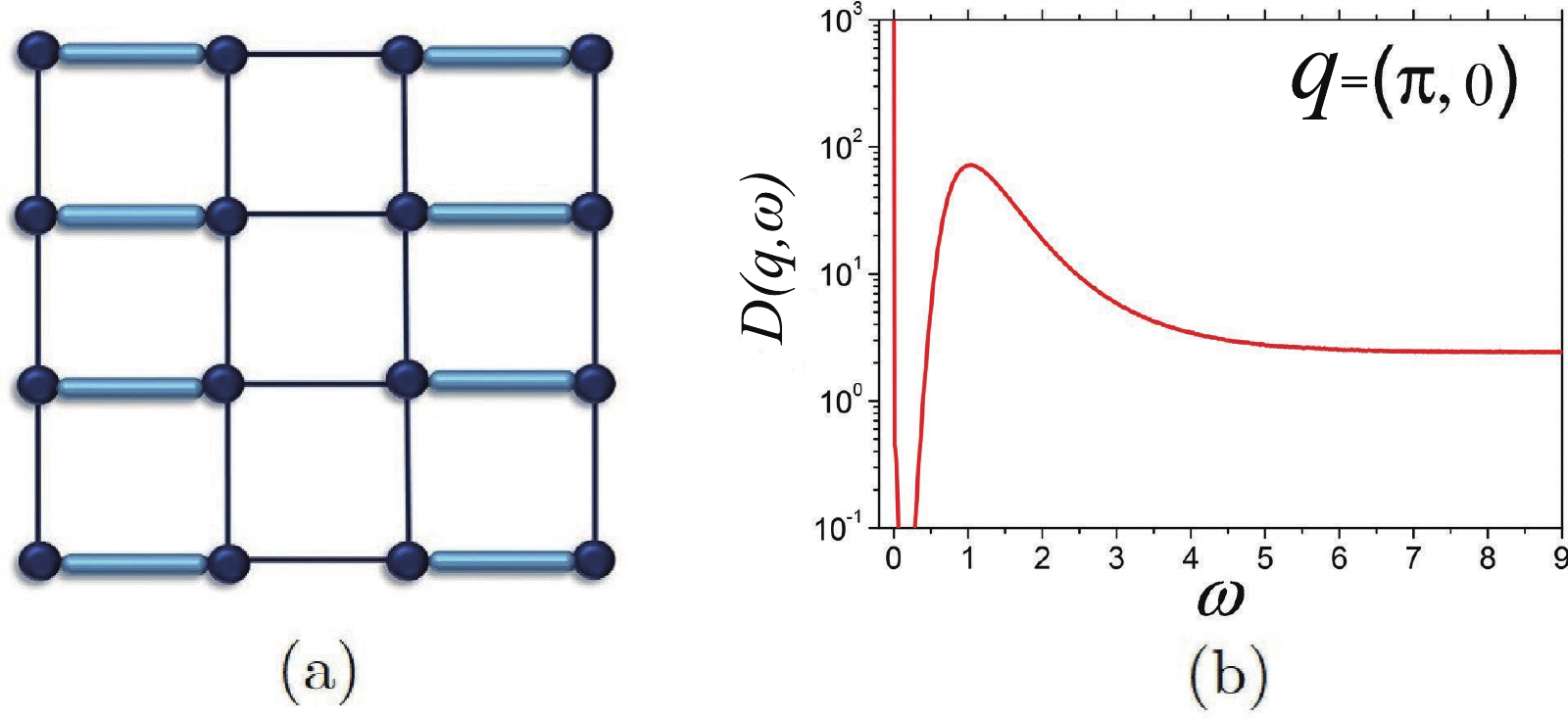}
\caption{To strengthen the VBS order, a pinning field is added as shown in (a), where the black bonds represent the original hopping $t$ in Eq.~\eqref{eq:eq1} while the blue bonds represent the enhanced hopping $t'= 1.05t$.  (b) shows the obtained spectrum at $(\pi,0)$ as a line cut in Fig.~\ref{fig:fig5} (d) in a log-scale, it is clear that inside the VBS phase, with the help of small pinning field, a Bragg peak ($\delta$-function) at $\omega=0$ plus a small gap and weight above it are revealed.}
\label{fig:fig6}
\end{figure}

As one moves into the U1D phase at $J=2.0<J_c$ (Fig.~\ref{fig:fig5} (b)), the gapless dimer spectra persist.  Again the spectral gaps at the high symmetry points are due to the finite size effect. The interesting observation here is that the continua are very broad, extending all the way from $\omega=0$ to $\omega=8J$, i.e., beyond the upper boundary of the non-interacting spectra. This points to the importance of higher order continua of multi-spinon excitations due to the strong interaction effect mediated by the U(1) gauge field fluctuations. It is also interesting to notice that at low energy, $\omega/J \le 2$, the spectral weight bear similar distribution with that of the non-interacting one, in particular, the weight is  greatly reducing as one approaches momentum $(0,0)$, this is related with the fact that $D^{x}_{(0,0)}$ is the energy-momentum tensor with larger scaling dimensions, as pointed out in Sec.~\ref{sec:iiic}. Also, the scaling dimension of $D^{x}_{(\pi,\pi)}$ could help with distinguishing the emergent symmetry of $\PSU(16)$ or $\PSU(8)\times\PSU(8)\times\mathbb{Z}_4$ in the U1D phase, given larger system sizes could be simulated.

Near and slightly above the critical point at $J=3.1$, as shown in Fig.~\ref{fig:fig5} (c), broad and prominent continuous spectra can still be observed and there are gapless spectra at $(\pi,0)$. The gapless excitation close to $(\pi,0)$ are the critical fluctuations associated with the QED$_3$-GN transition. With larger system sizes and lower temperature in the future QMC studies, one will be able to measure the anomalous dimension exponent $\eta$ from the momentum and frequency dependence of such critical fluctuation, and could compare with the predictions of QED$_3$-GN transitions from the recent perturbative RG calculations~\cite{JGracey2018,Ihrig2018,Zerf2019,Dupuis2019}.

Inside the VBS phase, the dimer spectra are gapped due to the $(\pi,0)$ or $(0,\pi)$ translational symmetry breaking. However, since the dimer order parameter will contribute a Bragg peak at $(\pi,0)$ and $\omega=0$, the analytic continuation is notoriously difficult for finding such not-smoothed spectra, i.e., one delta peak at $\omega=0$ followed with a gap and then continua above it. To solve this problem, we add a small pinning field to strengthen the VBS order~\cite{Assaad2013} in the simulation of Fig.~\ref{fig:fig5} (d). The pinning fields are added according to the pattern of Fig.~\ref{fig:fig6} (a), with $t'=1.05t$, and the simulation results are consistent with the expectation, in that, in Fig.~\ref{fig:fig5} (d), the spectra looked gapped at low energy, however, a fixed momentum cut at $(\pi,0)$, as shown in Fig.~\ref{fig:fig6} (b), indeed reveals that there is a Bragg peak at $\omega=0$ and a continuous spectra beyond a gap due to the break of discrete symmetry in VBS phase.

\section{Conclusions and discussion}
\label{sec:v}
In this work, we have performed both numerical and analytical analyses of the dynamics of a model realizing the compact QED$_3$ at large fermion flavor ($N_f=8$). As mentioned in the introduction, the question of U(1) gauge field coupled to fermionic matter field at (2+1)D is of high interests to both condensed matter and high-energy physics communities. The U1D phase is a realization of the algebraic quantum spin liquid in which Dirac spinons dynamically coupled to the emergent U(1) gauge field. The transition of U1D-VBS is the deconfinement-to-confinement transition in the QED$_3$ setting and is also the transition from symmetric quantum spin liquid to the symmetry-breaking phase that several potential quantum spin liquid compounds could have already realized by tuning the doping concentration, pressure and magnetic field~\cite{WeiqiangYu2017,Takagi2018}. The dynamical information of the U1D phase and U1D-VBS transition revealed here -- the continua in spin and dimer spectra and their field theoretical interpretations in the emergent symmetries and conserved currents -- provide the first piece of concrete evidence of the aforementioned exotic physical phenomena.

Looking forward, better algorithm in QMC simulations would certainly be desirable to access larger system sizes and lower temperatures. In particular, the critical properties of the U1D-VBS transition, that of the QED$_3$-GN types, have already been discussed in the high-order perturbative RG calculations~\cite{JGracey2018,Ihrig2018,Zerf2019,Dupuis2019}, but the system sizes in this work is too small to extract accurate values of the critical exponents. Further developments, in terms of algorithm improvement and more focus close to the QED$_3$-GN critical points, are on-going.

From analytical perspective, calculation of the spectra with the fluctuations of the U(1) gauge fields included would be very useful, similar as the analysis in the Ref.~\cite{PhysRevB.98.174421}, the dynamical signature of the strongly correlated systems of fractionalized spinons and their coupling effects with the emergent gauge field could be revealed and provide clearer guidance for future numerical simulations and eventually to experiments.

\section*{ACKNOWLEDGMENTS}
We thank Yang Qi, Lukas Janssen, Michael Scherer, Joseph Maciejko, John Gracey, Cenke Xu, Chong Wang, Yin-Chen He and Liujun Zou for helpful discussions. WW and ZYM acknowledge the supports from the Ministry of Science and Technology of China through the National Key Research and Development Program (Grant No. 2016YFA0300502), the Strategic Priority Research Program of the Chinese Academy of Sciences (Grant No. XDB28000000), and the National Science Foundation of China (Grant No. 11421092, 11574359 and 11674370). X. Y. X. is thankful for the support of Hong Kong Research Grants Council (HKRGC) through C6026-16W, 16324216 and 16307117. We thank the Center for Quantum Simulation Sciences at Institute of Physics, Chinese Academy of Sciences, and the Tianhe-1A platform at the National Supercomputer Center in Tianjin for their technical support and generous allocation of CPU time.

\bibliography{U1GaugeField}

\begin{thebibliography}{68}%
\makeatletter
\providecommand \@ifxundefined [1]{%
 \@ifx{#1\undefined}
}%
\providecommand \@ifnum [1]{%
 \ifnum #1\expandafter \@firstoftwo
 \else \expandafter \@secondoftwo
 \fi
}%
\providecommand \@ifx [1]{%
 \ifx #1\expandafter \@firstoftwo
 \else \expandafter \@secondoftwo
 \fi
}%
\providecommand \natexlab [1]{#1}%
\providecommand \enquote  [1]{``#1''}%
\providecommand \bibnamefont  [1]{#1}%
\providecommand \bibfnamefont [1]{#1}%
\providecommand \citenamefont [1]{#1}%
\providecommand \href@noop [0]{\@secondoftwo}%
\providecommand \href [0]{\begingroup \@sanitize@url \@href}%
\providecommand \@href[1]{\@@startlink{#1}\@@href}%
\providecommand \@@href[1]{\endgroup#1\@@endlink}%
\providecommand \@sanitize@url [0]{\catcode `\\12\catcode `\$12\catcode
  `\&12\catcode `\#12\catcode `\^12\catcode `\_12\catcode `\%12\relax}%
\providecommand \@@startlink[1]{}%
\providecommand \@@endlink[0]{}%
\providecommand \url  [0]{\begingroup\@sanitize@url \@url }%
\providecommand \@url [1]{\endgroup\@href {#1}{\urlprefix }}%
\providecommand \urlprefix  [0]{URL }%
\providecommand \Eprint [0]{\href }%
\providecommand \doibase [0]{http://dx.doi.org/}%
\providecommand \selectlanguage [0]{\@gobble}%
\providecommand \bibinfo  [0]{\@secondoftwo}%
\providecommand \bibfield  [0]{\@secondoftwo}%
\providecommand \translation [1]{[#1]}%
\providecommand \BibitemOpen [0]{}%
\providecommand \bibitemStop [0]{}%
\providecommand \bibitemNoStop [0]{.\EOS\space}%
\providecommand \EOS [0]{\spacefactor3000\relax}%
\providecommand \BibitemShut  [1]{\csname bibitem#1\endcsname}%
\let\auto@bib@innerbib\@empty
\bibitem [{\citenamefont {Xu}\ \emph {et~al.}(2019)\citenamefont {Xu},
  \citenamefont {Qi}, \citenamefont {Zhang}, \citenamefont {Assaad},
  \citenamefont {Xu},\ and\ \citenamefont {Meng}}]{Xiao2018Monte}%
  \BibitemOpen
  \bibfield  {author} {\bibinfo {author} {\bibfnamefont {Xiao~Yan}\
  \bibnamefont {Xu}}, \bibinfo {author} {\bibfnamefont {Yang}\ \bibnamefont
  {Qi}}, \bibinfo {author} {\bibfnamefont {Long}\ \bibnamefont {Zhang}},
  \bibinfo {author} {\bibfnamefont {Fakher~F.}\ \bibnamefont {Assaad}},
  \bibinfo {author} {\bibfnamefont {Cenke}\ \bibnamefont {Xu}}, \ and\ \bibinfo
  {author} {\bibfnamefont {Zi~Yang}\ \bibnamefont {Meng}},\ }\bibfield  {title}
  {\enquote {\bibinfo {title} {Monte carlo study of lattice compact quantum
  electrodynamics with fermionic matter: The parent state of quantum phases},}\
  }\href {\doibase 10.1103/PhysRevX.9.021022} {\bibfield  {journal} {\bibinfo
  {journal} {Phys. Rev. X}\ }\textbf {\bibinfo {volume} {9}},\ \bibinfo {pages}
  {021022} (\bibinfo {year} {2019})}\BibitemShut {NoStop}%
\bibitem [{\citenamefont {Fiebig}\ and\ \citenamefont
  {Woloshyn}(1990)}]{PhysRevD.42.3520}%
  \BibitemOpen
  \bibfield  {author} {\bibinfo {author} {\bibfnamefont {H.~R.}\ \bibnamefont
  {Fiebig}}\ and\ \bibinfo {author} {\bibfnamefont {R.~M.}\ \bibnamefont
  {Woloshyn}},\ }\bibfield  {title} {\enquote {\bibinfo {title} {{Monopoles and
  chiral-symmetry breaking in three-dimensional lattice QED}},}\ }\href
  {\doibase 10.1103/PhysRevD.42.3520} {\bibfield  {journal} {\bibinfo
  {journal} {Phys. Rev. D}\ }\textbf {\bibinfo {volume} {42}},\ \bibinfo
  {pages} {3520--3523} (\bibinfo {year} {1990})}\BibitemShut {NoStop}%
\bibitem [{\citenamefont {Armour}\ \emph {et~al.}(2011)\citenamefont {Armour},
  \citenamefont {Hands}, \citenamefont {Kogut}, \citenamefont {Lucini},
  \citenamefont {Strouthos},\ and\ \citenamefont
  {Vranas}}]{PhysRevD.84.014502}%
  \BibitemOpen
  \bibfield  {author} {\bibinfo {author} {\bibfnamefont {Wesley}\ \bibnamefont
  {Armour}}, \bibinfo {author} {\bibfnamefont {Simon}\ \bibnamefont {Hands}},
  \bibinfo {author} {\bibfnamefont {John~B.}\ \bibnamefont {Kogut}}, \bibinfo
  {author} {\bibfnamefont {Biagio}\ \bibnamefont {Lucini}}, \bibinfo {author}
  {\bibfnamefont {Costas}\ \bibnamefont {Strouthos}}, \ and\ \bibinfo {author}
  {\bibfnamefont {Pavlos}\ \bibnamefont {Vranas}},\ }\bibfield  {title}
  {\enquote {\bibinfo {title} {{Magnetic monopole plasma phase in $(2+1)d$
  compact quantum electrodynamics with fermionic matter}},}\ }\href {\doibase
  10.1103/PhysRevD.84.014502} {\bibfield  {journal} {\bibinfo  {journal} {Phys.
  Rev. D}\ }\textbf {\bibinfo {volume} {84}},\ \bibinfo {pages} {014502}
  (\bibinfo {year} {2011})}\BibitemShut {NoStop}%
\bibitem [{\citenamefont {Karthik}\ and\ \citenamefont
  {Narayanan}(2016)}]{PhysRevD.94.065026}%
  \BibitemOpen
  \bibfield  {author} {\bibinfo {author} {\bibfnamefont {Nikhil}\ \bibnamefont
  {Karthik}}\ and\ \bibinfo {author} {\bibfnamefont {Rajamani}\ \bibnamefont
  {Narayanan}},\ }\bibfield  {title} {\enquote {\bibinfo {title} {{Scale
  invariance of parity-invariant three-dimensional QED}},}\ }\href {\doibase
  10.1103/PhysRevD.94.065026} {\bibfield  {journal} {\bibinfo  {journal} {Phys.
  Rev. D}\ }\textbf {\bibinfo {volume} {94}},\ \bibinfo {pages} {065026}
  (\bibinfo {year} {2016})}\BibitemShut {NoStop}%
\bibitem [{\citenamefont {Karthik}\ and\ \citenamefont
  {Narayanan}(2018{\natexlab{a}})}]{PhysRevD.97.054510}%
  \BibitemOpen
  \bibfield  {author} {\bibinfo {author} {\bibfnamefont {Nikhil}\ \bibnamefont
  {Karthik}}\ and\ \bibinfo {author} {\bibfnamefont {Rajamani}\ \bibnamefont
  {Narayanan}},\ }\bibfield  {title} {\enquote {\bibinfo {title}
  {{Scale-invariance and scale-breaking in parity-invariant three-dimensional
  QCD}},}\ }\href {\doibase 10.1103/PhysRevD.97.054510} {\bibfield  {journal}
  {\bibinfo  {journal} {Phys. Rev. D}\ }\textbf {\bibinfo {volume} {97}},\
  \bibinfo {pages} {054510} (\bibinfo {year} {2018}{\natexlab{a}})}\BibitemShut
  {NoStop}%
\bibitem [{\citenamefont {Karthik}\ and\ \citenamefont
  {Narayanan}(2018{\natexlab{b}})}]{PhysRevLett.121.041602}%
  \BibitemOpen
  \bibfield  {author} {\bibinfo {author} {\bibfnamefont {Nikhil}\ \bibnamefont
  {Karthik}}\ and\ \bibinfo {author} {\bibfnamefont {Rajamani}\ \bibnamefont
  {Narayanan}},\ }\bibfield  {title} {\enquote {\bibinfo {title} {{Parity
  Anomaly Cancellation in Three-Dimensional QED with a Single Massless Dirac
  Fermion}},}\ }\href {\doibase 10.1103/PhysRevLett.121.041602} {\bibfield
  {journal} {\bibinfo  {journal} {Phys. Rev. Lett.}\ }\textbf {\bibinfo
  {volume} {121}},\ \bibinfo {pages} {041602} (\bibinfo {year}
  {2018}{\natexlab{b}})}\BibitemShut {NoStop}%
\bibitem [{\citenamefont {Braun}\ \emph {et~al.}(2014)\citenamefont {Braun},
  \citenamefont {Gies}, \citenamefont {Janssen},\ and\ \citenamefont
  {Roscher}}]{PhysRevD.90.036002}%
  \BibitemOpen
  \bibfield  {author} {\bibinfo {author} {\bibfnamefont {Jens}\ \bibnamefont
  {Braun}}, \bibinfo {author} {\bibfnamefont {Holger}\ \bibnamefont {Gies}},
  \bibinfo {author} {\bibfnamefont {Lukas}\ \bibnamefont {Janssen}}, \ and\
  \bibinfo {author} {\bibfnamefont {Dietrich}\ \bibnamefont {Roscher}},\
  }\bibfield  {title} {\enquote {\bibinfo {title} {{Phase structure of
  many-flavor ${\mathrm{QED}}_{3}$}},}\ }\href {\doibase
  10.1103/PhysRevD.90.036002} {\bibfield  {journal} {\bibinfo  {journal} {Phys.
  Rev. D}\ }\textbf {\bibinfo {volume} {90}},\ \bibinfo {pages} {036002}
  (\bibinfo {year} {2014})}\BibitemShut {NoStop}%
\bibitem [{\citenamefont {Kotikov}\ \emph {et~al.}(2016)\citenamefont
  {Kotikov}, \citenamefont {Shilin},\ and\ \citenamefont
  {Teber}}]{PhysRevD.94.056009}%
  \BibitemOpen
  \bibfield  {author} {\bibinfo {author} {\bibfnamefont {A.~V.}\ \bibnamefont
  {Kotikov}}, \bibinfo {author} {\bibfnamefont {V.~I.}\ \bibnamefont {Shilin}},
  \ and\ \bibinfo {author} {\bibfnamefont {S.}~\bibnamefont {Teber}},\
  }\bibfield  {title} {\enquote {\bibinfo {title} {{Critical behavior of
  ($2+1$)-dimensional QED: $1/{N}_{f}$ corrections in the Landau gauge}},}\
  }\href {\doibase 10.1103/PhysRevD.94.056009} {\bibfield  {journal} {\bibinfo
  {journal} {Phys. Rev. D}\ }\textbf {\bibinfo {volume} {94}},\ \bibinfo
  {pages} {056009} (\bibinfo {year} {2016})}\BibitemShut {NoStop}%
\bibitem [{\citenamefont {Affleck}\ and\ \citenamefont
  {Marston}(1988)}]{PhysRevB.37.3774}%
  \BibitemOpen
  \bibfield  {author} {\bibinfo {author} {\bibfnamefont {Ian}\ \bibnamefont
  {Affleck}}\ and\ \bibinfo {author} {\bibfnamefont {J.~Brad}\ \bibnamefont
  {Marston}},\ }\bibfield  {title} {\enquote {\bibinfo {title} {{Large-n limit
  of the Heisenberg-Hubbard model: Implications for high-${T}_{c}$
  superconductors}},}\ }\href {\doibase 10.1103/PhysRevB.37.3774} {\bibfield
  {journal} {\bibinfo  {journal} {Phys. Rev. B}\ }\textbf {\bibinfo {volume}
  {37}},\ \bibinfo {pages} {3774--3777} (\bibinfo {year} {1988})}\BibitemShut
  {NoStop}%
\bibitem [{\citenamefont {Marston}\ and\ \citenamefont
  {Affleck}(1989)}]{PhysRevB.39.11538}%
  \BibitemOpen
  \bibfield  {author} {\bibinfo {author} {\bibfnamefont {J.~Brad}\ \bibnamefont
  {Marston}}\ and\ \bibinfo {author} {\bibfnamefont {Ian}\ \bibnamefont
  {Affleck}},\ }\bibfield  {title} {\enquote {\bibinfo {title} {{Large-$n$
  limit of the Hubbard-Heisenberg model}},}\ }\href {\doibase
  10.1103/PhysRevB.39.11538} {\bibfield  {journal} {\bibinfo  {journal} {Phys.
  Rev. B}\ }\textbf {\bibinfo {volume} {39}},\ \bibinfo {pages} {11538--11558}
  (\bibinfo {year} {1989})}\BibitemShut {NoStop}%
\bibitem [{\citenamefont {Lee}\ \emph {et~al.}(1998)\citenamefont {Lee},
  \citenamefont {Nagaosa}, \citenamefont {Ng},\ and\ \citenamefont
  {Wen}}]{PhysRevB.57.6003}%
  \BibitemOpen
  \bibfield  {author} {\bibinfo {author} {\bibfnamefont {Patrick~A.}\
  \bibnamefont {Lee}}, \bibinfo {author} {\bibfnamefont {Naoto}\ \bibnamefont
  {Nagaosa}}, \bibinfo {author} {\bibfnamefont {Tai-Kai}\ \bibnamefont {Ng}}, \
  and\ \bibinfo {author} {\bibfnamefont {Xiao-Gang}\ \bibnamefont {Wen}},\
  }\bibfield  {title} {\enquote {\bibinfo {title} {{SU(2) formulation of the
  $t\ensuremath{-}J$ model: Application to underdoped cuprates}},}\ }\href
  {\doibase 10.1103/PhysRevB.57.6003} {\bibfield  {journal} {\bibinfo
  {journal} {Phys. Rev. B}\ }\textbf {\bibinfo {volume} {57}},\ \bibinfo
  {pages} {6003--6021} (\bibinfo {year} {1998})}\BibitemShut {NoStop}%
\bibitem [{\citenamefont {Rantner}\ and\ \citenamefont
  {Wen}(2001)}]{PhysRevLett.86.3871}%
  \BibitemOpen
  \bibfield  {author} {\bibinfo {author} {\bibfnamefont {Walter}\ \bibnamefont
  {Rantner}}\ and\ \bibinfo {author} {\bibfnamefont {Xiao-Gang}\ \bibnamefont
  {Wen}},\ }\bibfield  {title} {\enquote {\bibinfo {title} {{Electron Spectral
  Function and Algebraic Spin Liquid for the Normal State of Underdoped High
  ${T}_{c}$ Superconductors}},}\ }\href {\doibase 10.1103/PhysRevLett.86.3871}
  {\bibfield  {journal} {\bibinfo  {journal} {Phys. Rev. Lett.}\ }\textbf
  {\bibinfo {volume} {86}},\ \bibinfo {pages} {3871--3874} (\bibinfo {year}
  {2001})}\BibitemShut {NoStop}%
\bibitem [{\citenamefont {Rantner}\ and\ \citenamefont
  {Wen}()}]{PhysRevB.66.144501}%
  \BibitemOpen
  \bibfield  {author} {\bibinfo {author} {\bibfnamefont {Walter}\ \bibnamefont
  {Rantner}}\ and\ \bibinfo {author} {\bibfnamefont {Xiao-Gang}\ \bibnamefont
  {Wen}},\ }\bibfield  {title} {\enquote {\bibinfo {title} {{Spin correlations
  in the algebraic spin liquid: Implications for high-${T}_{c}$
  superconductors}},}\ }\href {\doibase 10.1103/PhysRevB.66.144501} {\bibinfo
  {journal} {Phys. Rev. B}\ ,\ \bibinfo {pages} {144501}}\BibitemShut {NoStop}%
\bibitem [{\citenamefont {Senthil}\ and\ \citenamefont
  {Fisher}(2000)}]{PhysRevB.62.7850}%
  \BibitemOpen
\bibfield  {journal} {  }\bibfield  {author} {\bibinfo {author} {\bibfnamefont
  {T.}~\bibnamefont {Senthil}}\ and\ \bibinfo {author} {\bibfnamefont {Matthew
  P.~A.}\ \bibnamefont {Fisher}},\ }\bibfield  {title} {\enquote {\bibinfo
  {title} {{${Z}_{2}$ gauge theory of electron fractionalization in strongly
  correlated systems}},}\ }\href {\doibase 10.1103/PhysRevB.62.7850} {\bibfield
   {journal} {\bibinfo  {journal} {Phys. Rev. B}\ }\textbf {\bibinfo {volume}
  {62}},\ \bibinfo {pages} {7850--7881} (\bibinfo {year} {2000})}\BibitemShut
  {NoStop}%
\bibitem [{\citenamefont {Senthil}\ and\ \citenamefont
  {Motrunich}(2002)}]{PhysRevB.66.205104}%
  \BibitemOpen
  \bibfield  {author} {\bibinfo {author} {\bibfnamefont {T.}~\bibnamefont
  {Senthil}}\ and\ \bibinfo {author} {\bibfnamefont {O.}~\bibnamefont
  {Motrunich}},\ }\bibfield  {title} {\enquote {\bibinfo {title} {Microscopic
  models for fractionalized phases in strongly correlated systems},}\ }\href
  {\doibase 10.1103/PhysRevB.66.205104} {\bibfield  {journal} {\bibinfo
  {journal} {Phys. Rev. B}\ }\textbf {\bibinfo {volume} {66}},\ \bibinfo
  {pages} {205104} (\bibinfo {year} {2002})}\BibitemShut {NoStop}%
\bibitem [{\citenamefont {Herbut}\ and\ \citenamefont
  {Seradjeh}(2003)}]{PhysRevLett.91.171601}%
  \BibitemOpen
  \bibfield  {author} {\bibinfo {author} {\bibfnamefont {Igor~F.}\ \bibnamefont
  {Herbut}}\ and\ \bibinfo {author} {\bibfnamefont {Babak~H.}\ \bibnamefont
  {Seradjeh}},\ }\bibfield  {title} {\enquote {\bibinfo {title} {{Permanent
  Confinement in the Compact ${\mathrm{Q}\mathrm{E}\mathrm{D}}_{3}$ with
  Fermionic Matter}},}\ }\href {\doibase 10.1103/PhysRevLett.91.171601}
  {\bibfield  {journal} {\bibinfo  {journal} {Phys. Rev. Lett.}\ }\textbf
  {\bibinfo {volume} {91}},\ \bibinfo {pages} {171601} (\bibinfo {year}
  {2003})}\BibitemShut {NoStop}%
\bibitem [{\citenamefont {Hermele}\ \emph
  {et~al.}(2004{\natexlab{a}})\citenamefont {Hermele}, \citenamefont {Senthil},
  \citenamefont {Fisher}, \citenamefont {Lee}, \citenamefont {Nagaosa},\ and\
  \citenamefont {Wen}}]{PhysRevB.70.214437}%
  \BibitemOpen
  \bibfield  {author} {\bibinfo {author} {\bibfnamefont {Michael}\ \bibnamefont
  {Hermele}}, \bibinfo {author} {\bibfnamefont {T.}~\bibnamefont {Senthil}},
  \bibinfo {author} {\bibfnamefont {Matthew P.~A.}\ \bibnamefont {Fisher}},
  \bibinfo {author} {\bibfnamefont {Patrick~A.}\ \bibnamefont {Lee}}, \bibinfo
  {author} {\bibfnamefont {Naoto}\ \bibnamefont {Nagaosa}}, \ and\ \bibinfo
  {author} {\bibfnamefont {Xiao-Gang}\ \bibnamefont {Wen}},\ }\bibfield
  {title} {\enquote {\bibinfo {title} {{Stability of $U(1)$ spin liquids in two
  dimensions}},}\ }\href {\doibase 10.1103/PhysRevB.70.214437} {\bibfield
  {journal} {\bibinfo  {journal} {Phys. Rev. B}\ }\textbf {\bibinfo {volume}
  {70}},\ \bibinfo {pages} {214437} (\bibinfo {year}
  {2004}{\natexlab{a}})}\BibitemShut {NoStop}%
\bibitem [{\citenamefont {Assaad}(2005)}]{PhysRevB.71.075103}%
  \BibitemOpen
  \bibfield  {author} {\bibinfo {author} {\bibfnamefont {F.~F.}\ \bibnamefont
  {Assaad}},\ }\bibfield  {title} {\enquote {\bibinfo {title} {{Phase diagram
  of the half-filled two-dimensional $\mathrm{SU}(N)$ Hubbard-Heisenberg model:
  A quantum Monte Carlo study}},}\ }\href {\doibase 10.1103/PhysRevB.71.075103}
  {\bibfield  {journal} {\bibinfo  {journal} {Phys. Rev. B}\ }\textbf {\bibinfo
  {volume} {71}},\ \bibinfo {pages} {075103} (\bibinfo {year}
  {2005})}\BibitemShut {NoStop}%
\bibitem [{\citenamefont {Hermele}\ \emph {et~al.}(2005)\citenamefont
  {Hermele}, \citenamefont {Senthil},\ and\ \citenamefont
  {Fisher}}]{PhysRevB.72.104404}%
  \BibitemOpen
  \bibfield  {author} {\bibinfo {author} {\bibfnamefont {Michael}\ \bibnamefont
  {Hermele}}, \bibinfo {author} {\bibfnamefont {T.}~\bibnamefont {Senthil}}, \
  and\ \bibinfo {author} {\bibfnamefont {Matthew P.~A.}\ \bibnamefont
  {Fisher}},\ }\bibfield  {title} {\enquote {\bibinfo {title} {Algebraic spin
  liquid as the mother of many competing orders},}\ }\href {\doibase
  10.1103/PhysRevB.72.104404} {\bibfield  {journal} {\bibinfo  {journal} {Phys.
  Rev. B}\ }\textbf {\bibinfo {volume} {72}},\ \bibinfo {pages} {104404}
  (\bibinfo {year} {2005})}\BibitemShut {NoStop}%
\bibitem [{\citenamefont {Hermele}\ \emph {et~al.}(2007)\citenamefont
  {Hermele}, \citenamefont {Senthil},\ and\ \citenamefont
  {Fisher}}]{PhysRevB.76.149906}%
  \BibitemOpen
  \bibfield  {author} {\bibinfo {author} {\bibfnamefont {Michael}\ \bibnamefont
  {Hermele}}, \bibinfo {author} {\bibfnamefont {T.}~\bibnamefont {Senthil}}, \
  and\ \bibinfo {author} {\bibfnamefont {Matthew P.~A.}\ \bibnamefont
  {Fisher}},\ }\bibfield  {title} {\enquote {\bibinfo {title} {{Erratum:
  Algebraic spin liquid as the mother of many competing orders [Phys. Rev. B
  72, 104404 (2005)]}},}\ }\href {\doibase 10.1103/PhysRevB.76.149906}
  {\bibfield  {journal} {\bibinfo  {journal} {Phys. Rev. B}\ }\textbf {\bibinfo
  {volume} {76}},\ \bibinfo {pages} {149906} (\bibinfo {year}
  {2007})}\BibitemShut {NoStop}%
\bibitem [{\citenamefont {Nogueira}\ and\ \citenamefont
  {Kleinert}(2008)}]{PhysRevB.77.045107}%
  \BibitemOpen
  \bibfield  {author} {\bibinfo {author} {\bibfnamefont {Flavio~S.}\
  \bibnamefont {Nogueira}}\ and\ \bibinfo {author} {\bibfnamefont {Hagen}\
  \bibnamefont {Kleinert}},\ }\bibfield  {title} {\enquote {\bibinfo {title}
  {{Compact quantum electrodynamics in $2+1$ dimensions and spinon
  deconfinement: A renormalization group analysis}},}\ }\href {\doibase
  10.1103/PhysRevB.77.045107} {\bibfield  {journal} {\bibinfo  {journal} {Phys.
  Rev. B}\ }\textbf {\bibinfo {volume} {77}},\ \bibinfo {pages} {045107}
  (\bibinfo {year} {2008})}\BibitemShut {NoStop}%
\bibitem [{\citenamefont {He}\ \emph {et~al.}(2017)\citenamefont {He},
  \citenamefont {Zaletel}, \citenamefont {Oshikawa},\ and\ \citenamefont
  {Pollmann}}]{PhysRevX.7.031020}%
  \BibitemOpen
  \bibfield  {author} {\bibinfo {author} {\bibfnamefont {Yin-Chen}\
  \bibnamefont {He}}, \bibinfo {author} {\bibfnamefont {Michael~P.}\
  \bibnamefont {Zaletel}}, \bibinfo {author} {\bibfnamefont {Masaki}\
  \bibnamefont {Oshikawa}}, \ and\ \bibinfo {author} {\bibfnamefont {Frank}\
  \bibnamefont {Pollmann}},\ }\bibfield  {title} {\enquote {\bibinfo {title}
  {{Signatures of Dirac Cones in a DMRG Study of the Kagome Heisenberg
  Model}},}\ }\href {\doibase 10.1103/PhysRevX.7.031020} {\bibfield  {journal}
  {\bibinfo  {journal} {Phys. Rev. X}\ }\textbf {\bibinfo {volume} {7}},\
  \bibinfo {pages} {031020} (\bibinfo {year} {2017})}\BibitemShut {NoStop}%
\bibitem [{\citenamefont {Assaad}\ and\ \citenamefont
  {Grover}(2016)}]{PhysRevX.6.041049}%
  \BibitemOpen
  \bibfield  {author} {\bibinfo {author} {\bibfnamefont {F.~F.}\ \bibnamefont
  {Assaad}}\ and\ \bibinfo {author} {\bibfnamefont {Tarun}\ \bibnamefont
  {Grover}},\ }\bibfield  {title} {\enquote {\bibinfo {title} {{Simple
  Fermionic Model of Deconfined Phases and Phase Transitions}},}\ }\href
  {\doibase 10.1103/PhysRevX.6.041049} {\bibfield  {journal} {\bibinfo
  {journal} {Phys. Rev. X}\ }\textbf {\bibinfo {volume} {6}},\ \bibinfo {pages}
  {041049} (\bibinfo {year} {2016})}\BibitemShut {NoStop}%
\bibitem [{\citenamefont {Gazit}\ \emph {et~al.}(2017)\citenamefont {Gazit},
  \citenamefont {Randeria},\ and\ \citenamefont
  {Vishwanath}}]{gazit2017emergent}%
  \BibitemOpen
  \bibfield  {author} {\bibinfo {author} {\bibfnamefont {Snir}\ \bibnamefont
  {Gazit}}, \bibinfo {author} {\bibfnamefont {Mohit}\ \bibnamefont {Randeria}},
  \ and\ \bibinfo {author} {\bibfnamefont {Ashvin}\ \bibnamefont
  {Vishwanath}},\ }\bibfield  {title} {\enquote {\bibinfo {title} {Emergent
  dirac fermions and broken symmetries in confined and deconfined phases of z2
  gauge theories},}\ }\href {https://doi.org/10.1038/nphys4028} {\bibfield
  {journal} {\bibinfo  {journal} {Nature Physics}\ }\textbf {\bibinfo {volume}
  {13}},\ \bibinfo {pages} {484 EP --} (\bibinfo {year} {2017})}\BibitemShut
  {NoStop}%
\bibitem [{\citenamefont {Gazit}\ \emph {et~al.}(2018)\citenamefont {Gazit},
  \citenamefont {Assaad}, \citenamefont {Sachdev}, \citenamefont {Vishwanath},\
  and\ \citenamefont {Wang}}]{gazit2018confinement}%
  \BibitemOpen
  \bibfield  {author} {\bibinfo {author} {\bibfnamefont {Snir}\ \bibnamefont
  {Gazit}}, \bibinfo {author} {\bibfnamefont {Fakher~F}\ \bibnamefont
  {Assaad}}, \bibinfo {author} {\bibfnamefont {Subir}\ \bibnamefont {Sachdev}},
  \bibinfo {author} {\bibfnamefont {Ashvin}\ \bibnamefont {Vishwanath}}, \ and\
  \bibinfo {author} {\bibfnamefont {Chong}\ \bibnamefont {Wang}},\ }\bibfield
  {title} {\enquote {\bibinfo {title} {{Confinement transition of Z$_2$ gauge
  theories coupled to massless fermions: Emergent quantum chromodynamics and SO
  (5) symmetry}},}\ }\href@noop {} {\bibfield  {journal} {\bibinfo  {journal}
  {Proceedings of the National Academy of Sciences}\ }\textbf {\bibinfo
  {volume} {115}},\ \bibinfo {pages} {E6987--E6995} (\bibinfo {year}
  {2018})}\BibitemShut {NoStop}%
\bibitem [{\citenamefont {Prosko}\ \emph {et~al.}(2017)\citenamefont {Prosko},
  \citenamefont {Lee},\ and\ \citenamefont {Maciejko}}]{PhysRevB.96.205104}%
  \BibitemOpen
  \bibfield  {author} {\bibinfo {author} {\bibfnamefont {Christian}\
  \bibnamefont {Prosko}}, \bibinfo {author} {\bibfnamefont {Shu-Ping}\
  \bibnamefont {Lee}}, \ and\ \bibinfo {author} {\bibfnamefont {Joseph}\
  \bibnamefont {Maciejko}},\ }\bibfield  {title} {\enquote {\bibinfo {title}
  {{Simple Z$_2$ lattice gauge theories at finite fermion density}},}\ }\href
  {\doibase 10.1103/PhysRevB.96.205104} {\bibfield  {journal} {\bibinfo
  {journal} {Phys. Rev. B}\ }\textbf {\bibinfo {volume} {96}},\ \bibinfo
  {pages} {205104} (\bibinfo {year} {2017})}\BibitemShut {NoStop}%
\bibitem [{\citenamefont {{Chen}}\ \emph {et~al.}(2019)\citenamefont {{Chen}},
  \citenamefont {{Xu}}, \citenamefont {{Qi}},\ and\ \citenamefont
  {{Meng}}}]{ChuangChen2019}%
  \BibitemOpen
  \bibfield  {author} {\bibinfo {author} {\bibfnamefont {Chuang}\ \bibnamefont
  {{Chen}}}, \bibinfo {author} {\bibfnamefont {Xiao~Yan}\ \bibnamefont {{Xu}}},
  \bibinfo {author} {\bibfnamefont {Yang}\ \bibnamefont {{Qi}}}, \ and\
  \bibinfo {author} {\bibfnamefont {Zi~Yang}\ \bibnamefont {{Meng}}},\
  }\bibfield  {title} {\enquote {\bibinfo {title} {{Metals' awkward cousin is
  found}},}\ }\href@noop {} {\bibfield  {journal} {\bibinfo  {journal} {arXiv
  e-prints}\ ,\ \bibinfo {eid} {arXiv:1904.12872}} (\bibinfo {year} {2019})},\
  \Eprint {http://arxiv.org/abs/1904.12872} {arXiv:1904.12872
  [cond-mat.str-el]} \BibitemShut {NoStop}%
\bibitem [{\citenamefont {Balents}\ \emph {et~al.}(2002)\citenamefont
  {Balents}, \citenamefont {Fisher},\ and\ \citenamefont
  {Girvin}}]{PhysRevB.65.224412}%
  \BibitemOpen
  \bibfield  {author} {\bibinfo {author} {\bibfnamefont {L.}~\bibnamefont
  {Balents}}, \bibinfo {author} {\bibfnamefont {M.~P.~A.}\ \bibnamefont
  {Fisher}}, \ and\ \bibinfo {author} {\bibfnamefont {S.~M.}\ \bibnamefont
  {Girvin}},\ }\bibfield  {title} {\enquote {\bibinfo {title}
  {{Fractionalization in an easy-axis Kagome antiferromagnet}},}\ }\href
  {\doibase 10.1103/PhysRevB.65.224412} {\bibfield  {journal} {\bibinfo
  {journal} {Phys. Rev. B}\ }\textbf {\bibinfo {volume} {65}},\ \bibinfo
  {pages} {224412} (\bibinfo {year} {2002})}\BibitemShut {NoStop}%
\bibitem [{\citenamefont {Hermele}\ \emph
  {et~al.}(2004{\natexlab{b}})\citenamefont {Hermele}, \citenamefont {Fisher},\
  and\ \citenamefont {Balents}}]{PhysRevB.69.064404}%
  \BibitemOpen
  \bibfield  {author} {\bibinfo {author} {\bibfnamefont {Michael}\ \bibnamefont
  {Hermele}}, \bibinfo {author} {\bibfnamefont {Matthew P.~A.}\ \bibnamefont
  {Fisher}}, \ and\ \bibinfo {author} {\bibfnamefont {Leon}\ \bibnamefont
  {Balents}},\ }\bibfield  {title} {\enquote {\bibinfo {title} {{Pyrochlore
  photons: The $U(1)$ spin liquid in a $S=\frac{1}{2}$ three-dimensional
  frustrated magnet}},}\ }\href {\doibase 10.1103/PhysRevB.69.064404}
  {\bibfield  {journal} {\bibinfo  {journal} {Phys. Rev. B}\ }\textbf {\bibinfo
  {volume} {69}},\ \bibinfo {pages} {064404} (\bibinfo {year}
  {2004}{\natexlab{b}})}\BibitemShut {NoStop}%
\bibitem [{\citenamefont {Senthil}\ \emph {et~al.}(2004)\citenamefont
  {Senthil}, \citenamefont {Balents}, \citenamefont {Sachdev}, \citenamefont
  {Vishwanath},\ and\ \citenamefont {Fisher}}]{PhysRevB.70.144407}%
  \BibitemOpen
  \bibfield  {author} {\bibinfo {author} {\bibfnamefont {T.}~\bibnamefont
  {Senthil}}, \bibinfo {author} {\bibfnamefont {Leon}\ \bibnamefont {Balents}},
  \bibinfo {author} {\bibfnamefont {Subir}\ \bibnamefont {Sachdev}}, \bibinfo
  {author} {\bibfnamefont {Ashvin}\ \bibnamefont {Vishwanath}}, \ and\ \bibinfo
  {author} {\bibfnamefont {Matthew P.~A.}\ \bibnamefont {Fisher}},\ }\bibfield
  {title} {\enquote {\bibinfo {title} {{Quantum criticality beyond the
  Landau-Ginzburg-Wilson paradigm}},}\ }\href {\doibase
  10.1103/PhysRevB.70.144407} {\bibfield  {journal} {\bibinfo  {journal} {Phys.
  Rev. B}\ }\textbf {\bibinfo {volume} {70}},\ \bibinfo {pages} {144407}
  (\bibinfo {year} {2004})}\BibitemShut {NoStop}%
\bibitem [{\citenamefont {Sandvik}(2007)}]{PhysRevLett.98.227202}%
  \BibitemOpen
  \bibfield  {author} {\bibinfo {author} {\bibfnamefont {Anders~W.}\
  \bibnamefont {Sandvik}},\ }\bibfield  {title} {\enquote {\bibinfo {title}
  {{Evidence for Deconfined Quantum Criticality in a Two-Dimensional Heisenberg
  Model with Four-Spin Interactions}},}\ }\href {\doibase
  10.1103/PhysRevLett.98.227202} {\bibfield  {journal} {\bibinfo  {journal}
  {Phys. Rev. Lett.}\ }\textbf {\bibinfo {volume} {98}},\ \bibinfo {pages}
  {227202} (\bibinfo {year} {2007})}\BibitemShut {NoStop}%
\bibitem [{\citenamefont {Qin}\ \emph {et~al.}(2017{\natexlab{a}})\citenamefont
  {Qin}, \citenamefont {He}, \citenamefont {You}, \citenamefont {Lu},
  \citenamefont {Sen}, \citenamefont {Sandvik}, \citenamefont {Xu},\ and\
  \citenamefont {Meng}}]{PhysRevX.7.031052}%
  \BibitemOpen
  \bibfield  {author} {\bibinfo {author} {\bibfnamefont {Yan~Qi}\ \bibnamefont
  {Qin}}, \bibinfo {author} {\bibfnamefont {Yuan-Yao}\ \bibnamefont {He}},
  \bibinfo {author} {\bibfnamefont {Yi-Zhuang}\ \bibnamefont {You}}, \bibinfo
  {author} {\bibfnamefont {Zhong-Yi}\ \bibnamefont {Lu}}, \bibinfo {author}
  {\bibfnamefont {Arnab}\ \bibnamefont {Sen}}, \bibinfo {author} {\bibfnamefont
  {Anders~W.}\ \bibnamefont {Sandvik}}, \bibinfo {author} {\bibfnamefont
  {Cenke}\ \bibnamefont {Xu}}, \ and\ \bibinfo {author} {\bibfnamefont
  {Zi~Yang}\ \bibnamefont {Meng}},\ }\bibfield  {title} {\enquote {\bibinfo
  {title} {{Duality between the Deconfined Quantum-Critical Point and the
  Bosonic Topological Transition}},}\ }\href {\doibase
  10.1103/PhysRevX.7.031052} {\bibfield  {journal} {\bibinfo  {journal} {Phys.
  Rev. X}\ }\textbf {\bibinfo {volume} {7}},\ \bibinfo {pages} {031052}
  (\bibinfo {year} {2017}{\natexlab{a}})}\BibitemShut {NoStop}%
\bibitem [{\citenamefont {Ma}\ \emph {et~al.}(2018)\citenamefont {Ma},
  \citenamefont {Sun}, \citenamefont {You}, \citenamefont {Xu}, \citenamefont
  {Vishwanath}, \citenamefont {Sandvik},\ and\ \citenamefont
  {Meng}}]{PhysRevB.98.174421}%
  \BibitemOpen
  \bibfield  {author} {\bibinfo {author} {\bibfnamefont {Nvsen}\ \bibnamefont
  {Ma}}, \bibinfo {author} {\bibfnamefont {Guang-Yu}\ \bibnamefont {Sun}},
  \bibinfo {author} {\bibfnamefont {Yi-Zhuang}\ \bibnamefont {You}}, \bibinfo
  {author} {\bibfnamefont {Cenke}\ \bibnamefont {Xu}}, \bibinfo {author}
  {\bibfnamefont {Ashvin}\ \bibnamefont {Vishwanath}}, \bibinfo {author}
  {\bibfnamefont {Anders~W.}\ \bibnamefont {Sandvik}}, \ and\ \bibinfo {author}
  {\bibfnamefont {Zi~Yang}\ \bibnamefont {Meng}},\ }\bibfield  {title}
  {\enquote {\bibinfo {title} {{Dynamical signature of fractionalization at a
  deconfined quantum critical point}},}\ }\href {\doibase
  10.1103/PhysRevB.98.174421} {\bibfield  {journal} {\bibinfo  {journal} {Phys.
  Rev. B}\ }\textbf {\bibinfo {volume} {98}},\ \bibinfo {pages} {174421}
  (\bibinfo {year} {2018})}\BibitemShut {NoStop}%
\bibitem [{\citenamefont {{Li}}\ \emph {et~al.}(2019)\citenamefont {{Li}},
  \citenamefont {{Jian}},\ and\ \citenamefont {{Yao}}}]{Li2019Deconfined}%
  \BibitemOpen
  \bibfield  {author} {\bibinfo {author} {\bibfnamefont {Zi-Xiang}\
  \bibnamefont {{Li}}}, \bibinfo {author} {\bibfnamefont {Shao-Kai}\
  \bibnamefont {{Jian}}}, \ and\ \bibinfo {author} {\bibfnamefont {Hong}\
  \bibnamefont {{Yao}}},\ }\bibfield  {title} {\enquote {\bibinfo {title}
  {{Deconfined quantum criticality and emergent SO(5) symmetry in fermionic
  systems}},}\ }\href@noop {} {\bibfield  {journal} {\bibinfo  {journal} {arXiv
  e-prints}\ ,\ \bibinfo {eid} {arXiv:1904.10975}} (\bibinfo {year} {2019})},\
  \Eprint {http://arxiv.org/abs/1904.10975} {arXiv:1904.10975
  [cond-mat.str-el]} \BibitemShut {NoStop}%
\bibitem [{\citenamefont {Zhou}\ \emph {et~al.}(2019)\citenamefont {Zhou},
  \citenamefont {Wu},\ and\ \citenamefont {Kou}}]{Zhou2019Quantum}%
  \BibitemOpen
  \bibfield  {author} {\bibinfo {author} {\bibfnamefont {Jiang}\ \bibnamefont
  {Zhou}}, \bibinfo {author} {\bibfnamefont {Ya-Jie}\ \bibnamefont {Wu}}, \
  and\ \bibinfo {author} {\bibfnamefont {Su-Peng}\ \bibnamefont {Kou}},\
  }\bibfield  {title} {\enquote {\bibinfo {title} {Quantum critical duality in
  two-dimensional dirac semimetals},}\ }\href {\doibase
  10.1088/1674-1056/28/1/017402} {\bibfield  {journal} {\bibinfo  {journal}
  {Chinese Physics B}\ }\textbf {\bibinfo {volume} {28}},\ \bibinfo {pages}
  {017402} (\bibinfo {year} {2019})}\BibitemShut {NoStop}%
\bibitem [{\citenamefont {Nandkishore}\ \emph {et~al.}(2012)\citenamefont
  {Nandkishore}, \citenamefont {Metlitski},\ and\ \citenamefont
  {Senthil}}]{PhysRevB.86.045128}%
  \BibitemOpen
  \bibfield  {author} {\bibinfo {author} {\bibfnamefont {Rahul}\ \bibnamefont
  {Nandkishore}}, \bibinfo {author} {\bibfnamefont {Max~A.}\ \bibnamefont
  {Metlitski}}, \ and\ \bibinfo {author} {\bibfnamefont {T.}~\bibnamefont
  {Senthil}},\ }\bibfield  {title} {\enquote {\bibinfo {title} {{Orthogonal
  metals: The simplest non-Fermi liquids}},}\ }\href {\doibase
  10.1103/PhysRevB.86.045128} {\bibfield  {journal} {\bibinfo  {journal} {Phys.
  Rev. B}\ }\textbf {\bibinfo {volume} {86}},\ \bibinfo {pages} {045128}
  (\bibinfo {year} {2012})}\BibitemShut {NoStop}%
\bibitem [{\citenamefont {Polyakov}(1977)}]{polyakov1977quark}%
  \BibitemOpen
  \bibfield  {author} {\bibinfo {author} {\bibfnamefont {Alexander~M}\
  \bibnamefont {Polyakov}},\ }\bibfield  {title} {\enquote {\bibinfo {title}
  {{Quark confinement and topology of gauge theories}},}\ }\href@noop {}
  {\bibfield  {journal} {\bibinfo  {journal} {Nuclear Physics B}\ }\textbf
  {\bibinfo {volume} {120}},\ \bibinfo {pages} {429--458} (\bibinfo {year}
  {1977})}\BibitemShut {NoStop}%
\bibitem [{\citenamefont {Mandelstam}(1976)}]{mandelstam1976vortices}%
  \BibitemOpen
  \bibfield  {author} {\bibinfo {author} {\bibfnamefont {S}~\bibnamefont
  {Mandelstam}},\ }\bibfield  {title} {\enquote {\bibinfo {title} {{Vortices
  and quark confinement in nonabelian gauge theories, Phys}},}\ }\href@noop {}
  {\bibfield  {journal} {\bibinfo  {journal} {Rept}\ }\textbf {\bibinfo
  {volume} {23}},\ \bibinfo {pages} {245} (\bibinfo {year} {1976})}\BibitemShut
  {NoStop}%
\bibitem [{\citenamefont {Case}\ \emph {et~al.}(2004)\citenamefont {Case},
  \citenamefont {Seradjeh},\ and\ \citenamefont {Herbut}}]{case2004self}%
  \BibitemOpen
  \bibfield  {author} {\bibinfo {author} {\bibfnamefont {Matthew~J}\
  \bibnamefont {Case}}, \bibinfo {author} {\bibfnamefont {Babak~H}\
  \bibnamefont {Seradjeh}}, \ and\ \bibinfo {author} {\bibfnamefont {Igor~F}\
  \bibnamefont {Herbut}},\ }\bibfield  {title} {\enquote {\bibinfo {title}
  {{Self-consistent theory of compact QED3 with relativistic fermions}},}\
  }\href@noop {} {\bibfield  {journal} {\bibinfo  {journal} {Nuclear Physics
  B}\ }\textbf {\bibinfo {volume} {676}},\ \bibinfo {pages} {572--586}
  (\bibinfo {year} {2004})}\BibitemShut {NoStop}%
\bibitem [{\citenamefont {{Song}}\ \emph
  {et~al.}(2018{\natexlab{a}})\citenamefont {{Song}}, \citenamefont {{He}},
  \citenamefont {{Vishwanath}},\ and\ \citenamefont {{Wang}}}]{XYSong2018}%
  \BibitemOpen
  \bibfield  {author} {\bibinfo {author} {\bibfnamefont {Xue-Yang}\
  \bibnamefont {{Song}}}, \bibinfo {author} {\bibfnamefont {Yin-Chen}\
  \bibnamefont {{He}}}, \bibinfo {author} {\bibfnamefont {Ashvin}\ \bibnamefont
  {{Vishwanath}}}, \ and\ \bibinfo {author} {\bibfnamefont {Chong}\
  \bibnamefont {{Wang}}},\ }\bibfield  {title} {\enquote {\bibinfo {title}
  {{From spinon band topology to the symmetry quantum numbers of monopoles in
  Dirac spin liquids}},}\ }\href@noop {} {\bibfield  {journal} {\bibinfo
  {journal} {arXiv e-prints}\ ,\ \bibinfo {eid} {arXiv:1811.11182}} (\bibinfo
  {year} {2018}{\natexlab{a}})},\ \Eprint {http://arxiv.org/abs/1811.11182}
  {arXiv:1811.11182 [cond-mat.str-el]} \BibitemShut {NoStop}%
\bibitem [{\citenamefont {{Mithat}}(2008)}]{Unsal2008}%
  \BibitemOpen
  \bibfield  {author} {\bibinfo {author} {\bibfnamefont {Unsal}\ \bibnamefont
  {{Mithat}}},\ }\bibfield  {title} {\enquote {\bibinfo {title} {{Topological
  symmetry, spin liquids and CFT duals of Polyakov model with massless
  fermions}},}\ }\href@noop {} {\bibfield  {journal} {\bibinfo  {journal}
  {arXiv e-prints}\ ,\ \bibinfo {eid} {arXiv:0804.4664}} (\bibinfo {year}
  {2008})},\ \Eprint {http://arxiv.org/abs/0804.4664} {arXiv:0804.4664
  [cond-mat.str-el]} \BibitemShut {NoStop}%
\bibitem [{\citenamefont {Gracey}(2018)}]{JGracey2018}%
  \BibitemOpen
  \bibfield  {author} {\bibinfo {author} {\bibfnamefont {J.~A.}\ \bibnamefont
  {Gracey}},\ }\bibfield  {title} {\enquote {\bibinfo {title} {{Fermion
  bilinear operator critical exponents at $O\mathbf{(}1/{N}^{2}\mathbf{)}$ in
  the QED-Gross-Neveu universality class}},}\ }\href {\doibase
  10.1103/PhysRevD.98.085012} {\bibfield  {journal} {\bibinfo  {journal} {Phys.
  Rev. D}\ }\textbf {\bibinfo {volume} {98}},\ \bibinfo {pages} {085012}
  (\bibinfo {year} {2018})}\BibitemShut {NoStop}%
\bibitem [{\citenamefont {Ihrig}\ \emph {et~al.}(2018)\citenamefont {Ihrig},
  \citenamefont {Janssen}, \citenamefont {Mihaila},\ and\ \citenamefont
  {Scherer}}]{Ihrig2018}%
  \BibitemOpen
  \bibfield  {author} {\bibinfo {author} {\bibfnamefont {Bernhard}\
  \bibnamefont {Ihrig}}, \bibinfo {author} {\bibfnamefont {Lukas}\ \bibnamefont
  {Janssen}}, \bibinfo {author} {\bibfnamefont {Luminita~N.}\ \bibnamefont
  {Mihaila}}, \ and\ \bibinfo {author} {\bibfnamefont {Michael~M.}\
  \bibnamefont {Scherer}},\ }\bibfield  {title} {\enquote {\bibinfo {title}
  {{Deconfined criticality from the ${\mathrm{QED}}_{3}$-Gross-Neveu model at
  three loops}},}\ }\href {\doibase 10.1103/PhysRevB.98.115163} {\bibfield
  {journal} {\bibinfo  {journal} {Phys. Rev. B}\ }\textbf {\bibinfo {volume}
  {98}},\ \bibinfo {pages} {115163} (\bibinfo {year} {2018})}\BibitemShut
  {NoStop}%
\bibitem [{\citenamefont {{Zerf}}\ \emph {et~al.}(2019)\citenamefont {{Zerf}},
  \citenamefont {{Boyack}}, \citenamefont {{Marquard}}, \citenamefont
  {{Gracey}},\ and\ \citenamefont {{Maciejko}}}]{Zerf2019}%
  \BibitemOpen
  \bibfield  {author} {\bibinfo {author} {\bibfnamefont {Nikolai}\ \bibnamefont
  {{Zerf}}}, \bibinfo {author} {\bibfnamefont {Rufus}\ \bibnamefont
  {{Boyack}}}, \bibinfo {author} {\bibfnamefont {Peter}\ \bibnamefont
  {{Marquard}}}, \bibinfo {author} {\bibfnamefont {John~A.}\ \bibnamefont
  {{Gracey}}}, \ and\ \bibinfo {author} {\bibfnamefont {Joseph}\ \bibnamefont
  {{Maciejko}}},\ }\bibfield  {title} {\enquote {\bibinfo {title} {{Critical
  properties of the Neel-to-algebraic spin liquid transition}},}\ }\href@noop
  {} {\bibfield  {journal} {\bibinfo  {journal} {arXiv e-prints}\ ,\ \bibinfo
  {eid} {arXiv:1905.03719}} (\bibinfo {year} {2019})},\ \Eprint
  {http://arxiv.org/abs/1905.03719} {arXiv:1905.03719 [cond-mat.str-el]}
  \BibitemShut {NoStop}%
\bibitem [{\citenamefont {{Dupuis}}\ \emph {et~al.}(2019)\citenamefont
  {{Dupuis}}, \citenamefont {{Paranjape}},\ and\ \citenamefont
  {{Witczak-Krempa}}}]{Dupuis2019}%
  \BibitemOpen
  \bibfield  {author} {\bibinfo {author} {\bibfnamefont {{\'E}ric}\
  \bibnamefont {{Dupuis}}}, \bibinfo {author} {\bibfnamefont {Manu}\
  \bibnamefont {{Paranjape}}}, \ and\ \bibinfo {author} {\bibfnamefont
  {William}\ \bibnamefont {{Witczak-Krempa}}},\ }\bibfield  {title} {\enquote
  {\bibinfo {title} {{Transition from a Dirac spin liquid to an
  antiferromagnet: Monopoles in a QED$_3$-Gross-Neveu theory}},}\ }\href@noop
  {} {\bibfield  {journal} {\bibinfo  {journal} {arXiv e-prints}\ ,\ \bibinfo
  {eid} {arXiv:1905.02750}} (\bibinfo {year} {2019})},\ \Eprint
  {http://arxiv.org/abs/1905.02750} {arXiv:1905.02750 [cond-mat.str-el]}
  \BibitemShut {NoStop}%
\bibitem [{\citenamefont {Zerf}\ \emph {et~al.}(2018)\citenamefont {Zerf},
  \citenamefont {Marquard}, \citenamefont {Boyack},\ and\ \citenamefont
  {Maciejko}}]{PhysRevB.98.165125}%
  \BibitemOpen
  \bibfield  {author} {\bibinfo {author} {\bibfnamefont {Nikolai}\ \bibnamefont
  {Zerf}}, \bibinfo {author} {\bibfnamefont {Peter}\ \bibnamefont {Marquard}},
  \bibinfo {author} {\bibfnamefont {Rufus}\ \bibnamefont {Boyack}}, \ and\
  \bibinfo {author} {\bibfnamefont {Joseph}\ \bibnamefont {Maciejko}},\
  }\bibfield  {title} {\enquote {\bibinfo {title} {{Critical behavior of the
  ${\mathrm{QED}}_{3}$-Gross-Neveu-Yukawa model at four loops}},}\ }\href
  {\doibase 10.1103/PhysRevB.98.165125} {\bibfield  {journal} {\bibinfo
  {journal} {Phys. Rev. B}\ }\textbf {\bibinfo {volume} {98}},\ \bibinfo
  {pages} {165125} (\bibinfo {year} {2018})}\BibitemShut {NoStop}%
\bibitem [{\citenamefont {Boyack}\ \emph {et~al.}(2019)\citenamefont {Boyack},
  \citenamefont {Rayyan},\ and\ \citenamefont {Maciejko}}]{PhysRevB.99.195135}%
  \BibitemOpen
  \bibfield  {author} {\bibinfo {author} {\bibfnamefont {Rufus}\ \bibnamefont
  {Boyack}}, \bibinfo {author} {\bibfnamefont {Ahmed}\ \bibnamefont {Rayyan}},
  \ and\ \bibinfo {author} {\bibfnamefont {Joseph}\ \bibnamefont {Maciejko}},\
  }\bibfield  {title} {\enquote {\bibinfo {title} {{Deconfined criticality in
  the ${\mathrm{QED}}_{3}$ Gross-Neveu-Yukawa model: The $1/N$ expansion
  revisited}},}\ }\href {\doibase 10.1103/PhysRevB.99.195135} {\bibfield
  {journal} {\bibinfo  {journal} {Phys. Rev. B}\ }\textbf {\bibinfo {volume}
  {99}},\ \bibinfo {pages} {195135} (\bibinfo {year} {2019})}\BibitemShut
  {NoStop}%
\bibitem [{\citenamefont {Sandvik}(2016)}]{PhysRevE.94.063308}%
  \BibitemOpen
  \bibfield  {author} {\bibinfo {author} {\bibfnamefont {Anders~W.}\
  \bibnamefont {Sandvik}},\ }\bibfield  {title} {\enquote {\bibinfo {title}
  {{Constrained sampling method for analytic continuation}},}\ }\href {\doibase
  10.1103/PhysRevE.94.063308} {\bibfield  {journal} {\bibinfo  {journal} {Phys.
  Rev. E}\ }\textbf {\bibinfo {volume} {94}},\ \bibinfo {pages} {063308}
  (\bibinfo {year} {2016})}\BibitemShut {NoStop}%
\bibitem [{\citenamefont {Shao}\ \emph {et~al.}(2017)\citenamefont {Shao},
  \citenamefont {Qin}, \citenamefont {Capponi}, \citenamefont {Chesi},
  \citenamefont {Meng},\ and\ \citenamefont {Sandvik}}]{Shao2017b}%
  \BibitemOpen
  \bibfield  {author} {\bibinfo {author} {\bibfnamefont {Hui}\ \bibnamefont
  {Shao}}, \bibinfo {author} {\bibfnamefont {Yan~Qi}\ \bibnamefont {Qin}},
  \bibinfo {author} {\bibfnamefont {Sylvain}\ \bibnamefont {Capponi}}, \bibinfo
  {author} {\bibfnamefont {Stefano}\ \bibnamefont {Chesi}}, \bibinfo {author}
  {\bibfnamefont {Zi~Yang}\ \bibnamefont {Meng}}, \ and\ \bibinfo {author}
  {\bibfnamefont {Anders~W.}\ \bibnamefont {Sandvik}},\ }\bibfield  {title}
  {\enquote {\bibinfo {title} {{Nearly Deconfined Spinon Excitations in the
  Square-Lattice Spin-$1/2$ Heisenberg Antiferromagnet}},}\ }\href {\doibase
  10.1103/PhysRevX.7.041072} {\bibfield  {journal} {\bibinfo  {journal} {Phys.
  Rev. X}\ }\textbf {\bibinfo {volume} {7}},\ \bibinfo {pages} {041072}
  (\bibinfo {year} {2017})}\BibitemShut {NoStop}%
\bibitem [{\citenamefont {Huang}\ \emph {et~al.}(2018)\citenamefont {Huang},
  \citenamefont {Deng}, \citenamefont {Wan},\ and\ \citenamefont
  {Meng}}]{Huang2017}%
  \BibitemOpen
  \bibfield  {author} {\bibinfo {author} {\bibfnamefont {Chun-Jiong}\
  \bibnamefont {Huang}}, \bibinfo {author} {\bibfnamefont {Youjin}\
  \bibnamefont {Deng}}, \bibinfo {author} {\bibfnamefont {Yuan}\ \bibnamefont
  {Wan}}, \ and\ \bibinfo {author} {\bibfnamefont {Zi~Yang}\ \bibnamefont
  {Meng}},\ }\bibfield  {title} {\enquote {\bibinfo {title} {{Dynamics of
  Topological Excitations in a Model Quantum Spin Ice}},}\ }\href {\doibase
  10.1103/PhysRevLett.120.167202} {\bibfield  {journal} {\bibinfo  {journal}
  {Phys. Rev. Lett.}\ }\textbf {\bibinfo {volume} {120}},\ \bibinfo {pages}
  {167202} (\bibinfo {year} {2018})}\BibitemShut {NoStop}%
\bibitem [{\citenamefont {Qin}\ \emph {et~al.}(2017{\natexlab{b}})\citenamefont
  {Qin}, \citenamefont {Normand}, \citenamefont {Sandvik},\ and\ \citenamefont
  {Meng}}]{qin2017amplitude}%
  \BibitemOpen
  \bibfield  {author} {\bibinfo {author} {\bibfnamefont {Yan~Qi}\ \bibnamefont
  {Qin}}, \bibinfo {author} {\bibfnamefont {B.}~\bibnamefont {Normand}},
  \bibinfo {author} {\bibfnamefont {Anders~W.}\ \bibnamefont {Sandvik}}, \ and\
  \bibinfo {author} {\bibfnamefont {Zi~Yang}\ \bibnamefont {Meng}},\ }\bibfield
   {title} {\enquote {\bibinfo {title} {{Amplitude Mode in Three-Dimensional
  Dimerized Antiferromagnets}},}\ }\href {\doibase
  10.1103/PhysRevLett.118.147207} {\bibfield  {journal} {\bibinfo  {journal}
  {Phys. Rev. Lett.}\ }\textbf {\bibinfo {volume} {118}},\ \bibinfo {pages}
  {147207} (\bibinfo {year} {2017}{\natexlab{b}})}\BibitemShut {NoStop}%
\bibitem [{\citenamefont {Sun}\ \emph {et~al.}(2018)\citenamefont {Sun},
  \citenamefont {Wang}, \citenamefont {Fang}, \citenamefont {Qi}, \citenamefont
  {Cheng},\ and\ \citenamefont {Meng}}]{GYSun2018}%
  \BibitemOpen
  \bibfield  {author} {\bibinfo {author} {\bibfnamefont {Guang-Yu}\
  \bibnamefont {Sun}}, \bibinfo {author} {\bibfnamefont {Yan-Cheng}\
  \bibnamefont {Wang}}, \bibinfo {author} {\bibfnamefont {Chen}\ \bibnamefont
  {Fang}}, \bibinfo {author} {\bibfnamefont {Yang}\ \bibnamefont {Qi}},
  \bibinfo {author} {\bibfnamefont {Meng}\ \bibnamefont {Cheng}}, \ and\
  \bibinfo {author} {\bibfnamefont {Zi~Yang}\ \bibnamefont {Meng}},\ }\bibfield
   {title} {\enquote {\bibinfo {title} {{Dynamical Signature of Symmetry
  Fractionalization in Frustrated Magnets}},}\ }\href {\doibase
  10.1103/PhysRevLett.121.077201} {\bibfield  {journal} {\bibinfo  {journal}
  {Phys. Rev. Lett.}\ }\textbf {\bibinfo {volume} {121}},\ \bibinfo {pages}
  {077201} (\bibinfo {year} {2018})}\BibitemShut {NoStop}%
\bibitem [{\citenamefont {Feng}\ \emph {et~al.}(2017)\citenamefont {Feng},
  \citenamefont {Li}, \citenamefont {Meng}, \citenamefont {Yi}, \citenamefont
  {Wei}, \citenamefont {Zhang}, \citenamefont {Wang}, \citenamefont {Jiang},
  \citenamefont {Liu}, \citenamefont {Li}, \citenamefont {Liu}, \citenamefont
  {Luo}, \citenamefont {Li}, \citenamefont {Zheng}, \citenamefont {Meng},
  \citenamefont {Mei},\ and\ \citenamefont {Shi}}]{Feng2017}%
  \BibitemOpen
  \bibfield  {author} {\bibinfo {author} {\bibfnamefont {Zili}\ \bibnamefont
  {Feng}}, \bibinfo {author} {\bibfnamefont {Zheng}\ \bibnamefont {Li}},
  \bibinfo {author} {\bibfnamefont {Xin}\ \bibnamefont {Meng}}, \bibinfo
  {author} {\bibfnamefont {Wei}\ \bibnamefont {Yi}}, \bibinfo {author}
  {\bibfnamefont {Yuan}\ \bibnamefont {Wei}}, \bibinfo {author} {\bibfnamefont
  {Jun}\ \bibnamefont {Zhang}}, \bibinfo {author} {\bibfnamefont {Yan-Cheng}\
  \bibnamefont {Wang}}, \bibinfo {author} {\bibfnamefont {Wei}\ \bibnamefont
  {Jiang}}, \bibinfo {author} {\bibfnamefont {Zheng}\ \bibnamefont {Liu}},
  \bibinfo {author} {\bibfnamefont {Shiyan}\ \bibnamefont {Li}}, \bibinfo
  {author} {\bibfnamefont {Feng}\ \bibnamefont {Liu}}, \bibinfo {author}
  {\bibfnamefont {Jianlin}\ \bibnamefont {Luo}}, \bibinfo {author}
  {\bibfnamefont {Shiliang}\ \bibnamefont {Li}}, \bibinfo {author}
  {\bibfnamefont {Guo-qing}\ \bibnamefont {Zheng}}, \bibinfo {author}
  {\bibfnamefont {Zi~Yang}\ \bibnamefont {Meng}}, \bibinfo {author}
  {\bibfnamefont {Jia-Wei}\ \bibnamefont {Mei}}, \ and\ \bibinfo {author}
  {\bibfnamefont {Youguo}\ \bibnamefont {Shi}},\ }\bibfield  {title} {\enquote
  {\bibinfo {title} {{Gapped Spin-1/2 Spinon Excitations in a New Kagome
  Quantum Spin Liquid Compound Cu$_3$Zn(OH)$_6$FBr}},}\ }\href {\doibase
  10.1088/0256-307X/34/7/077502} {\bibfield  {journal} {\bibinfo  {journal}
  {Chinese Physics Letters}\ }\textbf {\bibinfo {volume} {34}},\ \bibinfo {eid}
  {077502} (\bibinfo {year} {2017})}\BibitemShut {NoStop}%
\bibitem [{\citenamefont {{Wei}}\ \emph {et~al.}(2017)\citenamefont {{Wei}},
  \citenamefont {{Feng}}, \citenamefont {{Hu}}, \citenamefont {{Lohstroh}},
  \citenamefont {{dela Cruz}}, \citenamefont {{Yi}}, \citenamefont {{Ding}},
  \citenamefont {{Zhang}}, \citenamefont {{Tan}},\ and\ \citenamefont
  {{Shu}}}]{YuanWei2019}%
  \BibitemOpen
  \bibfield  {author} {\bibinfo {author} {\bibfnamefont {Yuan}\ \bibnamefont
  {{Wei}}}, \bibinfo {author} {\bibfnamefont {Zili}\ \bibnamefont {{Feng}}},
  \bibinfo {author} {\bibfnamefont {D.~H.}\ \bibnamefont {{Hu}}}, \bibinfo
  {author} {\bibfnamefont {Wiebke}\ \bibnamefont {{Lohstroh}}}, \bibinfo
  {author} {\bibfnamefont {Clarina}\ \bibnamefont {{dela Cruz}}}, \bibinfo
  {author} {\bibfnamefont {Wei}\ \bibnamefont {{Yi}}}, \bibinfo {author}
  {\bibfnamefont {Z.~F.}\ \bibnamefont {{Ding}}}, \bibinfo {author}
  {\bibfnamefont {J.}~\bibnamefont {{Zhang}}}, \bibinfo {author} {\bibfnamefont
  {Cheng}\ \bibnamefont {{Tan}}}, \ and\ \bibinfo {author} {\bibfnamefont
  {Lei}\ \bibnamefont {{Shu}}},\ }\bibfield  {title} {\enquote {\bibinfo
  {title} {{Evidence for a Z$_2$ topological ordered quantum spin liquid in a
  kagome-lattice antiferromagnet}},}\ }\href@noop {} {\bibfield  {journal}
  {\bibinfo  {journal} {arXiv e-prints}\ ,\ \bibinfo {eid} {arXiv:1710.02991}}
  (\bibinfo {year} {2017})},\ \Eprint {http://arxiv.org/abs/1710.02991}
  {arXiv:1710.02991 [cond-mat.str-el]} \BibitemShut {NoStop}%
\bibitem [{\citenamefont {Ma}\ \emph {et~al.}(2019)\citenamefont {Ma},
  \citenamefont {You},\ and\ \citenamefont {Meng}}]{PhysRevLett.122.175701}%
  \BibitemOpen
  \bibfield  {author} {\bibinfo {author} {\bibfnamefont {Nvsen}\ \bibnamefont
  {Ma}}, \bibinfo {author} {\bibfnamefont {Yi-Zhuang}\ \bibnamefont {You}}, \
  and\ \bibinfo {author} {\bibfnamefont {Zi~Yang}\ \bibnamefont {Meng}},\
  }\bibfield  {title} {\enquote {\bibinfo {title} {{Role of Noether's Theorem
  at the Deconfined Quantum Critical Point}},}\ }\href {\doibase
  10.1103/PhysRevLett.122.175701} {\bibfield  {journal} {\bibinfo  {journal}
  {Phys. Rev. Lett.}\ }\textbf {\bibinfo {volume} {122}},\ \bibinfo {pages}
  {175701} (\bibinfo {year} {2019})}\BibitemShut {NoStop}%
\bibitem [{\citenamefont {{Lee}}\ \emph {et~al.}(2019)\citenamefont {{Lee}},
  \citenamefont {{You}}, \citenamefont {{Sachdev}},\ and\ \citenamefont
  {{Vishwanath}}}]{JYLee2019}%
  \BibitemOpen
  \bibfield  {author} {\bibinfo {author} {\bibfnamefont {Jong~Yeon}\
  \bibnamefont {{Lee}}}, \bibinfo {author} {\bibfnamefont {Yi-Zhuang}\
  \bibnamefont {{You}}}, \bibinfo {author} {\bibfnamefont {Subir}\ \bibnamefont
  {{Sachdev}}}, \ and\ \bibinfo {author} {\bibfnamefont {Ashvin}\ \bibnamefont
  {{Vishwanath}}},\ }\bibfield  {title} {\enquote {\bibinfo {title}
  {{Signatures of a Deconfined Phase Transition on the Shastry-Sutherland
  Lattice: Applications to Quantum Critical SrCu$_2$(BO$_3$)$_2$}},}\
  }\href@noop {} {\bibfield  {journal} {\bibinfo  {journal} {arXiv e-prints}\
  ,\ \bibinfo {eid} {arXiv:1904.07266}} (\bibinfo {year} {2019})},\ \Eprint
  {http://arxiv.org/abs/1904.07266} {arXiv:1904.07266 [cond-mat.str-el]}
  \BibitemShut {NoStop}%
\bibitem [{\citenamefont {{Huang}}\ \emph {et~al.}(2019)\citenamefont
  {{Huang}}, \citenamefont {{Lu}}, \citenamefont {{You}}, \citenamefont
  {{Meng}},\ and\ \citenamefont {{Xiang}}}]{Huang2019}%
  \BibitemOpen
  \bibfield  {author} {\bibinfo {author} {\bibfnamefont {Rui-Zhen}\
  \bibnamefont {{Huang}}}, \bibinfo {author} {\bibfnamefont {Da-Chuan}\
  \bibnamefont {{Lu}}}, \bibinfo {author} {\bibfnamefont {Yi-Zhuang}\
  \bibnamefont {{You}}}, \bibinfo {author} {\bibfnamefont {Zi~Yang}\
  \bibnamefont {{Meng}}}, \ and\ \bibinfo {author} {\bibfnamefont {Tao}\
  \bibnamefont {{Xiang}}},\ }\bibfield  {title} {\enquote {\bibinfo {title}
  {{Emergent Symmetry and Conserved Current at a One Dimensional Incarnation of
  Deconfined Quantum Critical Point}},}\ }\href@noop {} {\bibfield  {journal}
  {\bibinfo  {journal} {arXiv e-prints}\ ,\ \bibinfo {eid} {arXiv:1904.00021}}
  (\bibinfo {year} {2019})},\ \Eprint {http://arxiv.org/abs/1904.00021}
  {arXiv:1904.00021 [cond-mat.str-el]} \BibitemShut {NoStop}%
\bibitem [{\citenamefont {{Xu}}\ \emph {et~al.}(2019)\citenamefont {{Xu}},
  \citenamefont {{Liu}}, \citenamefont {{Pan}}, \citenamefont {{Qi}},
  \citenamefont {{Sun}},\ and\ \citenamefont {{Meng}}}]{XYXu2019}%
  \BibitemOpen
  \bibfield  {author} {\bibinfo {author} {\bibfnamefont {Xiao~Yan}\
  \bibnamefont {{Xu}}}, \bibinfo {author} {\bibfnamefont {Zi~Hong}\
  \bibnamefont {{Liu}}}, \bibinfo {author} {\bibfnamefont {Gaopei}\
  \bibnamefont {{Pan}}}, \bibinfo {author} {\bibfnamefont {Yang}\ \bibnamefont
  {{Qi}}}, \bibinfo {author} {\bibfnamefont {Kai}\ \bibnamefont {{Sun}}}, \
  and\ \bibinfo {author} {\bibfnamefont {Zi~Yang}\ \bibnamefont {{Meng}}},\
  }\bibfield  {title} {\enquote {\bibinfo {title} {{Revealing Fermionic Quantum
  Criticality from New Monte Carlo Techniques}},}\ }\href@noop {} {\bibfield
  {journal} {\bibinfo  {journal} {arXiv e-prints}\ ,\ \bibinfo {eid}
  {arXiv:1904.07355}} (\bibinfo {year} {2019})},\ \Eprint
  {http://arxiv.org/abs/1904.07355} {arXiv:1904.07355 [cond-mat.str-el]}
  \BibitemShut {NoStop}%
\bibitem [{\citenamefont {{Song}}\ \emph
  {et~al.}(2018{\natexlab{b}})\citenamefont {{Song}}, \citenamefont {{Wang}},
  \citenamefont {{Vishwanath}},\ and\ \citenamefont {{He}}}]{XYSong2018b}%
  \BibitemOpen
  \bibfield  {author} {\bibinfo {author} {\bibfnamefont {Xue-Yang}\
  \bibnamefont {{Song}}}, \bibinfo {author} {\bibfnamefont {Chong}\
  \bibnamefont {{Wang}}}, \bibinfo {author} {\bibfnamefont {Ashvin}\
  \bibnamefont {{Vishwanath}}}, \ and\ \bibinfo {author} {\bibfnamefont
  {Yin-Chen}\ \bibnamefont {{He}}},\ }\bibfield  {title} {\enquote {\bibinfo
  {title} {{Unifying Description of Competing Orders in Two Dimensional Quantum
  Magnets}},}\ }\href@noop {} {\bibfield  {journal} {\bibinfo  {journal} {arXiv
  e-prints}\ ,\ \bibinfo {eid} {arXiv:1811.11186}} (\bibinfo {year}
  {2018}{\natexlab{b}})},\ \Eprint {http://arxiv.org/abs/1811.11186}
  {arXiv:1811.11186 [cond-mat.str-el]} \BibitemShut {NoStop}%
\bibitem [{\citenamefont {Borokhov}\ \emph {et~al.}(2002)\citenamefont
  {Borokhov}, \citenamefont {Kapustin},\ and\ \citenamefont
  {Wu}}]{Borokhov2002Topological}%
  \BibitemOpen
  \bibfield  {author} {\bibinfo {author} {\bibfnamefont {Vadim}\ \bibnamefont
  {Borokhov}}, \bibinfo {author} {\bibfnamefont {Anton}\ \bibnamefont
  {Kapustin}}, \ and\ \bibinfo {author} {\bibfnamefont {Xinkai}\ \bibnamefont
  {Wu}},\ }\bibfield  {title} {\enquote {\bibinfo {title} {Topological disorder
  operators in three-dimensional conformal field theory},}\ }\href {\doibase
  10.1088/1126-6708/2002/11/049} {\bibfield  {journal} {\bibinfo  {journal}
  {Journal of High Energy Physics}\ }\textbf {\bibinfo {volume} {2002}},\
  \bibinfo {pages} {049--049} (\bibinfo {year} {2002})}\BibitemShut {NoStop}%
\bibitem [{\citenamefont {{Dyer}}\ \emph {et~al.}(2013)\citenamefont {{Dyer}},
  \citenamefont {{Mezei}},\ and\ \citenamefont {{Pufu}}}]{Dyer2013Monopole}%
  \BibitemOpen
  \bibfield  {author} {\bibinfo {author} {\bibfnamefont {Ethan}\ \bibnamefont
  {{Dyer}}}, \bibinfo {author} {\bibfnamefont {M{\'a}rk}\ \bibnamefont
  {{Mezei}}}, \ and\ \bibinfo {author} {\bibfnamefont {Silviu~S.}\ \bibnamefont
  {{Pufu}}},\ }\bibfield  {title} {\enquote {\bibinfo {title} {{Monopole
  Taxonomy in Three-Dimensional Conformal Field Theories}},}\ }\href@noop {}
  {\bibfield  {journal} {\bibinfo  {journal} {arXiv e-prints}\ ,\ \bibinfo
  {eid} {arXiv:1309.1160}} (\bibinfo {year} {2013})},\ \Eprint
  {http://arxiv.org/abs/1309.1160} {arXiv:1309.1160 [hep-th]} \BibitemShut
  {NoStop}%
\bibitem [{\citenamefont {{Pufu}}\ and\ \citenamefont
  {{Sachdev}}(2013)}]{Pufu2013Monopoles}%
  \BibitemOpen
  \bibfield  {author} {\bibinfo {author} {\bibfnamefont {Silviu~S.}\
  \bibnamefont {{Pufu}}}\ and\ \bibinfo {author} {\bibfnamefont {Subir}\
  \bibnamefont {{Sachdev}}},\ }\bibfield  {title} {\enquote {\bibinfo {title}
  {{Monopoles in 2 + 1-dimensional conformal field theories with global U(1)
  symmetry}},}\ }\href {\doibase 10.1007/JHEP09(2013)127} {\bibfield  {journal}
  {\bibinfo  {journal} {Journal of High Energy Physics}\ }\textbf {\bibinfo
  {volume} {2013}},\ \bibinfo {eid} {127} (\bibinfo {year} {2013})},\ \Eprint
  {http://arxiv.org/abs/1303.3006} {arXiv:1303.3006 [hep-th]} \BibitemShut
  {NoStop}%
\bibitem [{Note1()}]{Note1}%
  \BibitemOpen
  \bibinfo {note} {Note that the notion of $N_f$ in the lattice model is
  differed from that in the field theory by a factor of 2}\BibitemShut
  {NoStop}%
\bibitem [{\citenamefont {Zhou}\ \emph {et~al.}(2018)\citenamefont {Zhou},
  \citenamefont {Wu},\ and\ \citenamefont {Wang}}]{Zhou2018Mott}%
  \BibitemOpen
  \bibfield  {author} {\bibinfo {author} {\bibfnamefont {Zhichao}\ \bibnamefont
  {Zhou}}, \bibinfo {author} {\bibfnamefont {Congjun}\ \bibnamefont {Wu}}, \
  and\ \bibinfo {author} {\bibfnamefont {Yu}~\bibnamefont {Wang}},\ }\bibfield
  {title} {\enquote {\bibinfo {title} {{Mott transition in the
  $\ensuremath{\pi}$-flux $SU(4)$ Hubbard model on a square lattice}},}\ }\href
  {\doibase 10.1103/PhysRevB.97.195122} {\bibfield  {journal} {\bibinfo
  {journal} {Phys. Rev. B}\ }\textbf {\bibinfo {volume} {97}},\ \bibinfo
  {pages} {195122} (\bibinfo {year} {2018})}\BibitemShut {NoStop}%
\bibitem [{\citenamefont {Kleinert}\ \emph {et~al.}(2002)\citenamefont
  {Kleinert}, \citenamefont {Nogueira},\ and\ \citenamefont
  {Sudb\o{}}}]{PhysRevLett.88.232001}%
  \BibitemOpen
  \bibfield  {author} {\bibinfo {author} {\bibfnamefont {Hagen}\ \bibnamefont
  {Kleinert}}, \bibinfo {author} {\bibfnamefont {Flavio~S.}\ \bibnamefont
  {Nogueira}}, \ and\ \bibinfo {author} {\bibfnamefont {Asle}\ \bibnamefont
  {Sudb\o{}}},\ }\bibfield  {title} {\enquote {\bibinfo {title} {{Deconfinement
  Transition in Three-Dimensional Compact $U(1)$ Gauge Theories Coupled to
  Matter Fields}},}\ }\href {\doibase 10.1103/PhysRevLett.88.232001} {\bibfield
   {journal} {\bibinfo  {journal} {Phys. Rev. Lett.}\ }\textbf {\bibinfo
  {volume} {88}},\ \bibinfo {pages} {232001} (\bibinfo {year}
  {2002})}\BibitemShut {NoStop}%
\bibitem [{\citenamefont {Assaad}\ and\ \citenamefont
  {Herbut}(2013)}]{Assaad2013}%
  \BibitemOpen
  \bibfield  {author} {\bibinfo {author} {\bibfnamefont {Fakher~F.}\
  \bibnamefont {Assaad}}\ and\ \bibinfo {author} {\bibfnamefont {Igor~F.}\
  \bibnamefont {Herbut}},\ }\bibfield  {title} {\enquote {\bibinfo {title}
  {{Pinning the Order: The Nature of Quantum Criticality in the Hubbard Model
  on Honeycomb Lattice}},}\ }\href {\doibase 10.1103/PhysRevX.3.031010}
  {\bibfield  {journal} {\bibinfo  {journal} {Phys. Rev. X}\ }\textbf {\bibinfo
  {volume} {3}},\ \bibinfo {pages} {031010} (\bibinfo {year}
  {2013})}\BibitemShut {NoStop}%
\bibitem [{\citenamefont {Zheng}\ \emph {et~al.}(2017)\citenamefont {Zheng},
  \citenamefont {Ran}, \citenamefont {Li}, \citenamefont {Wang}, \citenamefont
  {Wang}, \citenamefont {Liu}, \citenamefont {Liu}, \citenamefont {Normand},
  \citenamefont {Wen},\ and\ \citenamefont {Yu}}]{WeiqiangYu2017}%
  \BibitemOpen
  \bibfield  {author} {\bibinfo {author} {\bibfnamefont {Jiacheng}\
  \bibnamefont {Zheng}}, \bibinfo {author} {\bibfnamefont {Kejing}\
  \bibnamefont {Ran}}, \bibinfo {author} {\bibfnamefont {Tianrun}\ \bibnamefont
  {Li}}, \bibinfo {author} {\bibfnamefont {Jinghui}\ \bibnamefont {Wang}},
  \bibinfo {author} {\bibfnamefont {Pengshuai}\ \bibnamefont {Wang}}, \bibinfo
  {author} {\bibfnamefont {Bin}\ \bibnamefont {Liu}}, \bibinfo {author}
  {\bibfnamefont {Zhengxin}\ \bibnamefont {Liu}}, \bibinfo {author}
  {\bibfnamefont {B}~\bibnamefont {Normand}}, \bibinfo {author} {\bibfnamefont
  {Jinsheng}\ \bibnamefont {Wen}}, \ and\ \bibinfo {author} {\bibfnamefont
  {Weiqiang}\ \bibnamefont {Yu}},\ }\bibfield  {title} {\enquote {\bibinfo
  {title} {{Gapless Spin Excitations in the Field-Induced Quantum Spin Liquid
  Phase of $\alpha$-RuCl$_3$}},}\ }\href {\doibase
  10.1103/PhysRevLett.119.227208} {\bibfield  {journal} {\bibinfo  {journal}
  {Physical Review Letters}\ }\textbf {\bibinfo {volume} {119}} (\bibinfo
  {year} {2017}),\ 10.1103/PhysRevLett.119.227208}\BibitemShut {NoStop}%
\bibitem [{\citenamefont {Kitagawa}\ \emph {et~al.}(2018)\citenamefont
  {Kitagawa}, \citenamefont {Takayama}, \citenamefont {Matsumoto},
  \citenamefont {Kato}, \citenamefont {Takano}, \citenamefont {Kishimoto},
  \citenamefont {Bette}, \citenamefont {Dinnebier}, \citenamefont {Jackeli},\
  and\ \citenamefont {Takagi}}]{Takagi2018}%
  \BibitemOpen
  \bibfield  {author} {\bibinfo {author} {\bibfnamefont {K.}~\bibnamefont
  {Kitagawa}}, \bibinfo {author} {\bibfnamefont {T.}~\bibnamefont {Takayama}},
  \bibinfo {author} {\bibfnamefont {Y.}~\bibnamefont {Matsumoto}}, \bibinfo
  {author} {\bibfnamefont {A.}~\bibnamefont {Kato}}, \bibinfo {author}
  {\bibfnamefont {R.}~\bibnamefont {Takano}}, \bibinfo {author} {\bibfnamefont
  {Y.}~\bibnamefont {Kishimoto}}, \bibinfo {author} {\bibfnamefont
  {S.}~\bibnamefont {Bette}}, \bibinfo {author} {\bibfnamefont
  {R.}~\bibnamefont {Dinnebier}}, \bibinfo {author} {\bibfnamefont
  {G.}~\bibnamefont {Jackeli}}, \ and\ \bibinfo {author} {\bibfnamefont
  {H.}~\bibnamefont {Takagi}},\ }\bibfield  {title} {\enquote {\bibinfo {title}
  {{A spin--orbital-entangled quantum liquid on a honeycomb lattice}},}\ }\href
  {\doibase 10.1038/nature25482} {\bibfield  {journal} {\bibinfo  {journal}
  {Nature}\ }\textbf {\bibinfo {volume} {554}},\ \bibinfo {pages} {341}
  (\bibinfo {year} {2018})}\BibitemShut {NoStop}%
\end{thebibliography}%

\end{document}